# THEORETICAL STUDY OF THE SOFT OPTIC MODE DYNAMICS IN A RELAXOR FERROELECTRIC: THE EFFECT OF POLAR NANOREGIONS

E.Iolin[1] and J.Toulouse

Physics Department, Lehigh University, Bethlehem, PA 18015, USA

eui206@lehigh.edu

ABSTRACT

We propose a simple and solvable mean-field model of the scattering of transverse optic modes by Polarized Nano Regions (PNR) in the paraelectric phase of relaxor ferroelectrics. The PNR is assumed to be a ferroelectric sphere embedded in the host isotropic medium. The Lagrangian parameters are taken to be the same inside and outside the PNR, with exception of the soft-mode gap temperature dependence. The interaction of the Transverse (TO) with the longitudinal (LO) optic modes is taken into account but the latter is found to be important only at the surface of the PNR. Elementary excitations of the system are found to be of two types – Vortex (V) and Quasi Polar (QP). V excitations correspond to closed polarization lines or closed TO displacements while QP excitations contain open TO polarization lines, with an electric dipole if the net angular momentum *j*=1. LO waves are virtually excited only in the thin layer near the PNR surface and can be excluded from the boundary conditions without any consequence. The final boundary conditions include only V and QP TO wave amplitudes. Dynamical equations are solved and a phase diagram is predicted. The phase transition temperature is found to be lower for Vortex than for QP excitations. Therefore, the QP condensation (local phase transition) occurs before the Vortex one upon cooling. TO scattering by the PNR reveals strong long-wave resonances that can be shown to result from shallow localized and quasi-localized states and may be essential to understanding the waterfall observed in TO studies by inelastic neutron scattering.

## 1. INTRODUCTION

Over the past several decades, serious progress has been made in understanding the physical characteristics of the disordered ferroelectrics so-called relaxors (see for example reviews [1], [2]). These technically important materials (PMN, PZN, KTN and other) are characterized by chemical and structural disordering. Early on, Burns and Dacol [3] introduced the concept of Polarized Nano Regions (PNR) – small ordered (polarized) regions embedded into the host crystal - as the main feature of relaxors. Since then, the validity of this concept has been confirmed by numerous optical, diffuse and inelastic neutron scattering studies (see for example [4, 5, 6]).
An important observation was made by Gehring et al [6, 7] in their study of the Transverse Optic mode (soft-mode) dispersion in the relaxor PZN-8%PT in a temperature range where PNR's should certainly exist. These authors found that, while perfectly well defined TO peaks were observed in the neutron scattering spectra (constant Q-scan) for momenta $q>q_0$, these peaks disappeared for $q<q_0$ and the value of $1/q_0$ was comparable to the expected PNR size of a few tens of angstroms. In addition, the dispersion curves had the shape of a "waterfall". Different explanations of this phenomenon have been proposed [7 ÷ 9]. Gehring et al [8] attributed their results to a sharp step-like increase in the TO damping for $q<q_0$ (that could be described by an even Fermi function). However, no firm consensus has been reached as to the definite origin of the waterfall phenomenon. Another motivation for the present theoretical work are our own neutron scattering measurements of the transverse acoustic mode (TA) dispersion and damping in the relaxor $K_{1-x}Ta_xNbO_3$ (KTN) [10,11]. The shape of the TA peaks exhibit clear deviations from a standard Lorentzian form (Q-scans) at several TA momenta. These deviations appear to be more or less regular, not easily explained by experimental uncertainty and could be interpreted as evidence for resonant scattering of the TA by PNRs.

[1] Author to whom correspondence should be addressed. E-mail: eui206@lehigh.edu

It is now generally accepted that polarization correlations (PNR) appear at the so-called Burns temperature, $T_B$, upon cooling. It has also been argued [12] that the correlations appearing at $T=T_B$ are dynamic and become static at a significantly lower temperature, $T^*$, corresponding to a local distortion or local phase transition. $T^*$ is in fact the temperature at which elastic diffuse scattering appears in neutron spectra as well as the characteristic relaxor behavior or frequency dispersion of the dielectric constant. Consequently, the term *polar nanodomains* (PND) has been proposed to describe these polarization regions. So far, we do not have detail information about the shape of these regions, the nature of their boundaries (sharp or smooth), the value of the surface energy etc. In the present work (see also [13]), we consider PNRs in the temperature interval $T^*<T<T_B$, but nevertheless suppose that the relaxor can be approximated as a two-component system. We then describe theoretically the scattering of the transverse optic phonon (TO) by PNRs in the framework of the simplest possible model – a spherically symmetric PNR embedded in an isotropic medium. For simplicity, we assume that all Hamiltonian (Lagrangian) parameters are the same outside (index 1) and inside (index 2) the PNR, with exception of the temperature of the phase transition (Tc1, Tc2) and the energy gaps $\omega_0(1)$, $\omega_0(2)$ of the TOW soft-mode at zero momentum, q=0. We also assume that all parameters are constant. Mixing of transverse and longitudinal waves at the PNR surface is taken into account but we assume the same gap parameter for the LO and TO in order to avoid using higher derivatives in our equations. In the Lagrangian, we therefore include the interaction of the polarization with the electromagnetic field and take into account longitudinal Coulomb forces. However, we describe the system with a critical TO and non critical LO behavior because the LO does not usually soften in ferroelectrics and because Coulomb forces tend to raise its frequency at small momentum in ionic crystals, leading to a large energy gap $\Omega_p$ >> $\omega_0(1)$, $\omega_0(2)$ and only a moderate (non-critical) dependence on temperature. From electrodynamics, we find that elementary transverse excitations of the system can be divided into two categories – Vortex (V) and quasi polar (QP[2]). Vortex excitations correspond to closed or loop polarization lines of TO displacements and QP excitations contain open TO polarization lines (if the total angular momentum j=1). LO waves are virtually excited ($\Omega_p$>> $\omega_0(1)$, $\omega_0(2)$ ) but only in a thin layer near the PNR surface. In the limit $\Omega_p$->∞, we can analytically exclude them from the boundary conditions without restriction on the validity of the results. Hence, the final boundary conditions contain only the TO Vortex and QP amplitudes. The vortex boundary conditions are standard and correspond to the condition of continuous TO displacement and its derivative with respect to the radial coordinate at the PNR surface. Quasi-Polar excitations correspond to the condition of a continuous TO displacement but discontinuous derivative with respect to the radial coordinate at the PNR surface. This discontinuity of the derivative results in a near-field effect that vanishes for a large size PNR.

[2] Similar photon excitations are called as magnetic and electric in electrodynamics [16].

In the following parts of paper we formulate the model and provide its solution for the boundary conditions described above. We show that a local instability ("phase transition in PNR") occurs at a temperature $T_L<Tc_2$ where the mode frequency goes to zero locally, $\omega_L\approx 0$ (if $Tc_2>Tc_1$, $Tc_2$ is then the critical temperature for the case of a very large size PNR). $T_L$ is found to be lower for the case of a Vortex excitation than for a QP excitation. Therefore, the QP condensation (local phase transition) is expected to occur before the Vortex condensation upon cooling. Also, $T_L$ is found to be lower for a smaller size PNR. For $Tc_2<Tc_1$, we also find that local TO modes can exist, for which $\omega_L < \omega_0(1)$, due to reflection by the host medium. In the last part of our paper, we calculate the Vortex and QP scattering by the PNR and find strong resonance scattering at small TO momentum. All calculations are exact as we do not assume that the TO–TO interaction between PNR and host medium is small. A short discussion of the results and future perspective are presented in the conclusion section. Some mathematical derivations are detailed in the Appendix.

## 2. MODEL

### *2.1 MODEL DESCRIPTION.*

We consider the dynamics of a spherically symmetric PNR embedded in an isotropic medium. The total Lagrangian density $L$ contains the contributions of the TO and LO waves, $L_o$, the electrostatic field [16], $L_{em}$, and the interaction between electric field and polarization, $L_{int}$.

The total Lagrangian Λ is written as:

$$\Lambda = \int L d^3 r, \quad L = L_o + L_{em} + L_{\text{int}} \tag{1}$$

$$L_o = K_o - \Pi_o \quad \text{where } K_o = 1/2\rho_o (\partial\xi/\partial t)^2 \text{ and } \Pi_o = +(A_o/2) div(\xi)^2 + B_o \tilde{\xi}_{ik}^{\,2} + 1/2\rho_o \omega_0^2 \xi^2$$

$$\xi_{ik} = 1/2(\partial\xi_i/\partial x_k + \partial\xi_k/\partial x_i), \quad \tilde{\xi}_{ik} = \xi_{ik} - 1/3\delta_{ik}\xi_{qq} \tag{2}$$

$$L_{em} = \frac{1}{8\pi}(\nabla\varphi)^2 \tag{3}$$

$$L_{\text{int}} = -e^* \xi_\alpha \nabla_\alpha \varphi \tag{4}$$

Here $\boldsymbol{\xi}$ is the displacement in the optic mode, $\rho_o$ the corresponding density, $e^*$ the effective charge. $A_o$, and $B_o$ are parameters describing the optic mode contribution to the potential energy written in a similar form to that in elasticity theory [14] and $\varphi$ is the electric scalar potential. A similar Lagrangian was proposed by Hopfield [15] for the description of polariton dynamics. Our Lagrangian is different from that proposed by Hopfield in three essential aspects. First, we do not take into account polariton effects. Second, we take into account the space dispersion of optical modes (see $\Pi_o$ (1) first and second terms), and, third, we also take into account the softening of the optic mode in the last term of $\Pi_o$:

$$\omega_0^2(1) = a1^2(T - Tc_1), \quad \omega_0^2(2) = a2^2(T - Tc_2), \quad Tc_1 < Tc_2, \ \text{a1} < \text{a2} \tag{5}$$

A phase transition is expected to occur inside the PNR at a higher temperature than in the host medium, $T_{c1} < T_{c2}$, resulting in what we called earlier a polar nanodomain. We should also note that Hopfield used the expression $\boldsymbol{L_H}$ for the description of the interaction between electric field and polarization, $L_H = e^* \varphi \nabla_\alpha \xi_\alpha = L_{\text{int}} + e^* \nabla_\alpha (\xi_\alpha \varphi)$ Both expressions $\boldsymbol{L_H}$ and (4) lead to the same results for the case of infinitely large medium. We prefer apply expression (4) which is similar to the standard description of interaction between polarization and electric field and polarization [1] and more suitable for the boundary conditions formulation. Given the above Lagrangian, the dynamical equations are written in the standard form:

$$d / dt(\delta\Lambda / \delta(d\xi_\alpha(r) / dt)) = \delta\Lambda / \delta\xi_\alpha(r), \;\; \delta\Lambda / \delta\varphi(r) = 0 \qquad (6)$$

From Eq.2 we then get:

$$\begin{aligned} &\rho_0 \partial^2 \xi_\alpha / \partial t^2 = Ao * grad_\alpha div(\xi) + 2Bo * \nabla_k (\tilde{\xi}_{\alpha k}) - \\ &-\rho_0 \omega_o^2 \xi_\alpha - e^* \nabla_\alpha \varphi - Ao \int div(\xi) dS_\alpha - 2Bo * \int \tilde{\xi}_{\alpha k} dS_k \end{aligned} \qquad (7a)$$

and from Eqs.3 and 4:

$$-\nabla_\alpha (\nabla_\alpha \varphi - 4\pi e^* \xi_\alpha) + \int (\nabla_\alpha \varphi - 4\pi e^* \xi_\alpha) dS_\alpha = 0 \qquad (7b)$$

The electrodynamics boundary condition for the tangential component of the electric field is given by:

$$\nabla\varphi_1 - \mathbf{n}(\mathbf{n}\nabla\varphi_1) = \nabla\varphi_2 - \mathbf{n}(\mathbf{n}\nabla\varphi_2) \qquad (7c)$$

where $\boldsymbol{dS}$ is an element of the PNR surface and $\boldsymbol{n}$ is a unit vector normal to the surface. The terms containing $\boldsymbol{dS}$ (7a, b) lead to particular solutions satisfying the boundary conditions at the PNR interface. (7b, c) are the standard boundary conditions of electrodynamics. It is important to emphasize that all parameters are the same inside and outside the PNR with the exception of the gaps, $\omega_0(1)$ and $\omega_0(2)$, determined by the transition temperatures $T_{c1}$ and $T_{c2}$ respectively. We now derive separate independent solutions inside and outside the PNR in the form of plane waves with frequency $\omega$ and momentum $\boldsymbol{K_L}$, $\boldsymbol{K_T}$ for the optic longitudinal and transverse modes respectively. From Eqs.7a and 7b:

*Longitudinal Optic wave (LOW)*

$$\begin{aligned} &[\rho_o \omega^2 - AoK_L^2 - 4/3BoK_L^2 - \rho_o \omega_0^2]\xi_L - iK_L e^* \varphi = 0, \\ &iK_L \varphi - 4\pi e^* \xi_L = 0 \end{aligned} \qquad (8a)$$

Solving this system of two coupled equations, we get:

$$\omega^2 - AoK_L^2/\rho_o - 4/3BoK_L^2/\rho_o - \omega_0^2 - \Omega_p^2 = 0,$$

$$\text{with} \quad \Omega_p^2 = 4\pi(e^*)^2/\rho_O \qquad (8b)$$

Because the ion plasma frequency, $\boldsymbol{\Omega_p}$~15 - 75 meV >>$\omega_o$, is large, the softening of the LOW can be neglected.

*Transverse Optic wave (TOW)*

$$[\omega^2 - BoK_T^2/\rho_O - \omega_0^2]\xi_T = 0 \qquad (9)$$

For convenience, we write the parameter $\boldsymbol{A_0}$ and $\boldsymbol{B_0}$ in the LOW and TOW dispersion, Eqs. 8b and 9, in terms of the speeds $C_L$, $C_O$:

$$Ao \equiv \rho_O[C_L^2 - 4/3\,C_O^2],\quad Bo \equiv \rho_0 C_O^2,\ \xi \rightarrow \xi/\sqrt{\rho_0} \qquad (10)$$

We also take the density parameter, $\boldsymbol{\rho_0}$, to be equal to 1.

*2.2 MODEL SOLUTION (GENERAL).*

Because the effective potential energies inside or outside the PNR are independent of the coordinates (Eqs.1-4), the solution can be decomposed into independent radial and angular parts. The radial part can be written in terms of Bessel (Hankel) functions inside (outside) the PNR with argument $rK_T$ and $rK_L$ for the TOW and LOW respectively. The angular part is written in terms of spherical functions. We are especially interested in the dynamics of transverse excitations. A similar problem was considered in quantum electrodynamics [16]. Our case is similar but does not quite coincide with [16] because the longitudinal optic modes are important in our case while they are absent in [16], and because the optic mode energy spectrum contains a gap. For convenience, we utilize some results from Refs. [16], [17] and [18] concerning spherical vector functions. Because we are interested in scattering by a spherical PNR and the total angular momentum is conserved, it is convenient to use an angular momentum technique in the following. We describe the optic phonon polarization in terms of the vector spherical functions, $\chi_\mu(\alpha)$, $\mu$=0, ±1, $\alpha$=1,2,3 (x, y, z):

$$\chi_{-1} = \frac{1}{\sqrt{2}}\begin{pmatrix}1\\-i\\0\end{pmatrix},\quad \chi_0 = \begin{pmatrix}0\\0\\1\end{pmatrix},\quad \chi_1 = \frac{-1}{\sqrt{2}}\begin{pmatrix}1\\i\\0\end{pmatrix} \qquad (11)$$

The phonon wave field $Y_{jlM}$ contains angular and polarization ("spin") components in the *total* angular momentum $\boldsymbol{j}$ representation:

$$\boldsymbol{Y_{jlM}}(\mathbf{v},\alpha) = \sum_{\boldsymbol{m},\mu} \boldsymbol{C^{jM}_{lm1\mu}}\boldsymbol{Y_{lm}}(\mathbf{v})\chi_\mu(\alpha), \text{ with } \boldsymbol{m}+\mu=\boldsymbol{M} \text{ and } \mathbf{v}=\mathbf{k}/\boldsymbol{k},$$

$$\mathbf{v}^2 = 1,\quad \boldsymbol{l} = \boldsymbol{j}, \boldsymbol{j}\pm 1 \text{ if } \boldsymbol{j}\neq 0;\ \boldsymbol{l}=1 \text{ if } \boldsymbol{j}=0 \text{ and } \boldsymbol{M} = \boldsymbol{-j...j} \qquad (12)$$

where $\boldsymbol{k}$ is the momentum, $Y_{lm}$ is a standard spherical function [16-18] with angular momentum $l$ and its projection $m$ and $C^{jM}_{lm1\mu}$ is the Clebsch-Gordon coefficient [16]. We should emphasize that we are working here in the momentum (not-space) representation. Expression (12) can be written in the form of orthogonal normalized spherical vector functions (integration is done over the surface of a sphere of unit radius).

$$\mathbf{Y}_{jlM}(\mathbf{v}) = \sum_{m,\mu} C^{jM}_{lm1\mu} Y_{lm}(\mathbf{v})\chi_{\mu}, \quad \int \mathbf{Y}^{*}_{jlM}(\mathbf{v})\mathbf{Y}_{j'l'M'}(\mathbf{v})d\Omega = \delta_{jj'}\delta_{ll'}\delta_{MM'} \qquad (13)$$

We distinguish several solutions: one longitudinal, with spherical vector $Y^{(-1)}_{jM}(\mathbf{v})$ (superscript -1), directed along the momentum ***k***, and two transverse, $Y^{(0)}_{jM}(\mathbf{v})$ and $Y^{(1)}_{jM}(\mathbf{v})$ (superscript 0 and +1), with different parity). Several useful expressions concerning spherical wave functions [16 -19] are used in the paper:

$$\mathbf{Y}^{(-1)}_{jM}(\mathbf{v}) \equiv \mathbf{v}Y_{jM}(\mathbf{v}),\ \mathbf{Y}^{(0)}_{jM}(\mathbf{v}) \equiv \mathbf{Y}_{jjM}(\mathbf{v}),\ \mathbf{Y}^{(1)}_{jM}(\mathbf{v}) \equiv i[\mathbf{Y}_{jjM}(\mathbf{v}), \mathbf{v}],$$

$$\mathbf{Y}^{(0)}_{jM}(\mathbf{v})^{*}\mathbf{v}=0,\ \mathbf{Y}^{(0)}_{jM}(\mathbf{v})=-i\frac{[\mathbf{k}\nabla_{\mathbf{k}}]Y_{jM}(\mathbf{v})}{\sqrt{j(j+1)}},$$

$$\mathbf{Y}^{(1)}_{jM}(\mathbf{v}) * \mathbf{v} = 0,\ \mathbf{Y}^{(1)}_{jM}(\mathbf{v}) = \frac{k}{\sqrt{j(j+1)}}\nabla_{\mathbf{k}}Y_{jM}(\mathbf{v}), \qquad (14)$$

$$\mathbf{Y}^{(-1)}_{jM}(\mathbf{v}) = \sqrt{\frac{j}{2j+1}}\mathbf{Y}_{j,j-1,M}(\mathbf{v}) - \sqrt{\frac{j+1}{2j+1}}\mathbf{Y}_{j,j+1,M}(\mathbf{v}),$$

$$\mathbf{Y}^{(1)}_{jM}(\mathbf{v}) = \sqrt{\frac{j}{2j+1}}\mathbf{Y}_{j,j+1,M}(\mathbf{v}) + \sqrt{\frac{j+1}{2j+1}}\mathbf{Y}_{j,j-1,M}(\mathbf{v}) \qquad (15)$$

We note that the functions $\mathbf{Y}_{jM}^{(0)}(\mathbf{v})$, $\mathbf{Y}_{jM}^{(1)}(\mathbf{v})$ correspond to Vortex (V) and Quasi polar (QP) excitations respectively. A Vortex excitation is a manifold of closed lines of optic mode displacements (similar to the magnetic field lines in electrodynamics), and therefore does not possess an electric dipole moment. A QP TOW displacement (15) contains a spatial component corresponding to the spherical symmetric *S*-orbital state but also spin angular momentum, giving a non-zero electric dipole moment after integration over angles when *j*=1. We note also that only the longitudinal wave exists when the total angular momentum *j*=0.

Next, in order to fulfill the boundary conditions at the PNR surface, the phonon wave field must be expressed in real space. This can be done by means of an expansion of the plane wave in terms of spherical harmonics [16]:

$$\exp(i\mathbf{kr}) = \sum_{l,m} g_l(kr)Y^{*}_{lm}(\mathbf{k}/k)Y_{lm}(\mathbf{n}),\ \mathbf{n} = \mathbf{r}/r,\ \mathrm{n}^2 = 1,$$

$$g_l(kr)=(2\pi)^{3/2}i^l J_{l+1/2}(kr)/\sqrt{kr},$$

$$\int \mathbf{Y}_{jlM}(\mathbf{k}/k)\exp(i\mathbf{kr})d\Omega_k = g_l(kr)\mathbf{Y}_{jlm}(\mathbf{r}/r) \qquad (16)$$

$J_{l+1/2}$ is an ordinary Bessel function. As usual, Bessel functions are used in expression (16) inside PNR to ensure a finite TOW displacement at the PNR center, r=0; outside the PNR, we should use the first (second) order Hankel $H_{l+1/2}^{(1),(2)}$ function for the case of an “outward wave” (“inward wave”) instead of a Bessel function. According to our earlier notation, the wave vector outside (inside) the PNR is labeled as $k=K_T(1)$ ($k=K_T(2)$). In the real space representation, the wave fields corresponding to the $Y^{(0)}_{jM}(\mathbf{v})$, $Y^{(1)}_{jM}(\mathbf{v})$, *and* $Y^{(-1)}_{jM}(\mathbf{v})$ components are transformed into the expressions $\mathbf{F}^{(0)}_{jM}$, $\mathbf{F}^{(1)}_{jM}$, *and* $\mathbf{F}^{(-1)}_{jM}$, respectively:

$$\mathbf{F}_{jM}^{(0)} = g_j(kr)\mathbf{Y}_{jM}^{(0)}(\mathbf{n}) \equiv f_j^{(0)}(kr)\mathbf{Y}_{jM}^{(0)}(\mathbf{n}), \quad (17a)$$

$$\mathbf{F}_{jM}^{(1)} = [\frac{j}{2j+1}g_{j+1}(kr)+\frac{j+1}{2j+1}g_{j-1}(kr)]\mathbf{Y}_{jM}^{(1)}(\mathbf{n})+$$

$$\frac{\sqrt{j(j+1)}}{2j+1}[-g_{j+1}(kr)+g_{j-1}(kr)]\mathbf{Y}_{jM}^{(-1)}(\mathbf{n})\equiv$$

$$f_j^{(1,1)}(\mathrm{kr})\mathbf{Y}_{jM}^{(1)}(\mathbf{n})+f_j^{(1,-1)}(\mathrm{kr})\mathbf{Y}_{jM}^{(-1)}(\mathbf{n}) \quad (17b)$$

$$\mathbf{F}_{jM}^{(-1)} = [\frac{j}{2j+1}g_{j-1}(kr)+\frac{j+1}{2j+1}g_{j+1}(kr)]\mathbf{Y}_{jM}^{(-1)}(\mathbf{n})+$$

$$\frac{\sqrt{j(j+1)}}{2\mathrm{j}+1}[g_{j-1}(kr)-g_{j+1}(kr)]\mathbf{Y}_{jM}^{(1)}(\mathbf{n})\equiv$$

$$f_j^{(-1,-1)}(\mathrm{kr})\mathbf{Y}_{jM}^{(-1)}(\mathbf{n})+f_j^{(-1,1)}(\mathrm{kr})\mathbf{Y}_{jM}^{(1)}(\mathbf{n}) \quad (17c)$$

where we have introduced the coefficients $f^{(0)}_j$, $f^{(1,1)}_j$, $f^{(1,-1)}_j$, $f^{(-1,1)}_j$, $f^{(-1,-1)}$ for the expansion of the wave field components in terms of the orthogonal basis functions $\boldsymbol{Y}^{(0)}_{jM}(\boldsymbol{n})$, $\boldsymbol{Y}^{(1)}_{jM}(\boldsymbol{n})$, $\boldsymbol{Y}^{(-1)}_{jM}(\boldsymbol{n})$.

It is important to note that, when going from the momentum to the real-space representation, a transverse vortex (0) excitations converts to a transverse vortex excitation while a transverse quasi-polar (+1) converts to a combination of transverse (+1) and longitudinal (-1 ) excitations (17b). Hence, the index ±1 in the space representation and ±1 in the momentum representation do not have the same physical meaning. Polarization is normally defined with respect to $\boldsymbol{k}$, i.e. in the momentum representation. However, the boundary conditions are applied in real space, so that the conversion from momentum to real space representation is necessary. The main physical characteristic of the TOW (div($\boldsymbol{\xi}$)=0) is of course invariant with respect to the type of representation used. The functions $\boldsymbol{F}^{(0)}_{jM}$, $\boldsymbol{F}^{(+1)}_{jM}$ are transverse in the momentum representation ($\mathbf{F}^{(0)}_{jM}\,\mathbf{v}=0$, $\mathbf{F}^{(+1)}_{jM}\,\mathbf{v}=0$), i.e. with respect to k. In the space presentation however, the function $\boldsymbol{F}^{(0)}_{jM}$ is still transverse but this time with respect to $\boldsymbol{r}$ and the function $\boldsymbol{F}^{(+1)}_{jM}$ contains a *longitudinal* $\boldsymbol{Y}^{(-1)}_{jM}\boldsymbol{r}\neq 0$ term (17b). This term describes near field effects and is proportional to $1/r^2$ in the wave zone (kr>>l). At a large distance from the center of the PNR,

$$kr \to \infty, \; g_l(kr) \approx (2\pi)^{3/2}\sqrt{\frac{2}{\pi}}(i)^l \frac{\sin(kr-\frac{l\pi}{2})}{kr} \quad (18)$$

We refer to $\boldsymbol{F}^{(0)}_{jM}$ as the "Vortex" and to $\boldsymbol{F}^{(1)}_{jM}$ as the "quasi polar", QP, components of the TOW. In a QP $\boldsymbol{F}^{(1)}_{jM}$ type excitation, both transverse and longitudinal optical mode components will be present as a result of applying the boundary conditions (7a), because these BCs are stated for the tangential and normal components of the displacements and stresses, with respect to $\boldsymbol{r}$ and not $\boldsymbol{k}$. By contrast, the Vortex $\boldsymbol{F}^{(0)}_{jM}$ excitation will correspond to a strictly transverse excitation and will not be mixed with a LOW, since it only contains a $\boldsymbol{Y}^{0}_{jM}$ term, as seen in expression (17a). The general displacement $\boldsymbol{\xi}_L$ in the longitudinal optical wave (8b) can also be written as the gradient of some scalar "potential" $\Phi_o$ (19).

$$\xi_L = \nabla\Phi_o, \text{ with } \Phi_o = \mathrm{G}_\mathrm{j}(r)Y_{jM}(\vartheta,\phi) \quad (19)$$

Writing ξ in terms of the potential $\phi_o$ in Eqs. 7a and 7b, we get:

$$\Delta\Phi_o + (\omega^2 - \omega_0^2 - \Omega_p^2)/C_L^2\,\Phi_o = 0 \qquad (20a)$$

and

$$\frac{1}{r^2}\frac{\partial}{\partial r}\left(r^2\frac{\partial G_j(r)}{\partial r}\right) - \frac{j(j+1)}{r^2}G_j(r) + \frac{(\omega^2-\omega_0^2-\Omega_p^2)}{C_L^2}G_j(r) = 0 \quad (20b)$$

The LOW displacement (21) contains not only a component $\propto \boldsymbol{Y}^{(-1)}_{jM}(\boldsymbol{n})$ that is perpendicular to surface of the PNR but also a component $\propto \boldsymbol{Y}^{(1)}_{jM}(\boldsymbol{n})$ that is parallel to it ($\boldsymbol{n}=\boldsymbol{r}/r$, $\boldsymbol{n}^2=1$).

$$\xi_L = \frac{\partial G_j}{\partial r}\mathbf{n}Y_{jM} + G_j\nabla Y_{jM} = \frac{\partial G_j}{\partial r}\mathbf{Y}^{(-1)}_{jM}(\mathbf{n}) + G_j\frac{\sqrt{j(j+1)}}{r}\mathbf{Y}^{(1)}_{jM}(\mathbf{n}) \qquad (21)$$

The near field effects described by the term $\boldsymbol{Y}^{(1)}_{jM}(\boldsymbol{n})G_j/r$ in Eq.21 rapidly decay at a large distance $r$ from the PNR center $\sim 1/r^2$. Also, we are interested by the case when the ion plasma frequency in (8b), $\boldsymbol{\Omega_p}$, is very large, $\Omega_p \gg \omega$. Such a LOW will only be excited within a ***narrow layer*** of thickness δR at the PNR surface, $\delta R \sim 1/|2\pi K_L|$ (8b). For the case of a relaxor crystal, a typical PNR radius is $\boldsymbol{R}\sim 4 \div 15$ l.u. and δR~1 l.u. if $\Omega_p$=30meV and $C_L$=100 meV*l.u.

We now use the general solution of the model to describe the scattering of an incident transverse optic (TO) plane wave by a spherical PNR. Expanding the plane wave in terms of spherical harmonics [13], we obtain an expression for the partial wave $\boldsymbol{\sigma}^{(0)}_j$, $\boldsymbol{\sigma}^{(1)}_j$ and total, $\boldsymbol{\sigma}^{(0)}$, $\boldsymbol{\sigma}^{(1)}$ scattering cross-section:

$$\sigma^{(0,1)} = \sum_{j\geq 1}(2j+1)\sin(\delta_j^{(0,1)})^2\, 2\pi/k^2 \qquad (22)$$

where $\delta_j^{(0)}$ and $\delta_j^{(1)}$ are the phase shift for scattering of vortex and quasi-polar excitations respectively.

### *2.3 BOUNDARY CONDITIONS*

We begin with the simple case of Vortex excitations. The TOW displacement can be written as (17a), where $\boldsymbol{f_j}$ is a Bessel function inside the PNR and a Hankel function outside. The last term in Eq. (7a) requires that both the optical displacement and its derivative (strain) in the Vortex excitation be continuous through the surface

$$\xi = f_j(kr)\mathbf{Y}^{(0)}_{jM}(\mathbf{n}),\ \ \xi(\mathrm{r} = \mathrm{R} - \varepsilon) = \xi(\mathrm{r} = \mathrm{R} + \varepsilon),\ \ \varepsilon \to +0 \quad (23a)$$

$$\tilde{\xi}_{\alpha\beta} n_\beta = \frac{1}{2}\left(\frac{\partial \xi_\alpha}{\partial r_\beta} + \frac{\partial \xi_\beta}{\partial r_\alpha}\right) n_\beta = \frac{1}{2}\left(\frac{\partial f_j}{\partial r} - \frac{f_j}{r}\right) Y^{(0)}_{jM,\alpha},$$

$$\tilde{\xi}_{\alpha\beta} n_\beta(\mathrm{r} = \mathrm{R} - \varepsilon) = \tilde{\xi}_{\alpha\beta} n_\beta(\mathrm{r} = \mathrm{R} + \varepsilon) \qquad (23b)$$

We see that the vortex components are not mixed with quasi-polar excitations.
The vortex excitation boundary conditions can finally be ***exactly*** written in a compact form:

$$\Psi(g) = 0,\ \ \frac{\partial \Psi(g)}{\partial r} = 0 \qquad (24a)$$

where the auxiliary function Ψ(g) represents the difference between the TO amplitudes outside and inside the PNR:

$$\Psi(g) = A^{(0)}_{in} g_{j,in} + A^{(0)}_{\mathrm{out}} g_{j,out} - A^{(0)}(2) g_j \qquad (24b)$$

$A^{(0)}(2)$, $A^{(0)}_{in}$, $A^{(0)}_{out}$ are the Vortex amplitudes respectively inside the PNR, incident (in) and reflected (out) from the PNR surface on the outside.

The derivation of the boundary conditions is more complicated for the case of QP excitations (see Appendix). However, these conditions can be written in a compact form when the frequency of the longitudinal optical mode is very large. They can be expressed by means of the auxiliary functions $\Phi(G^{(1)})$ and $\Phi(G^{(-1)})$:

$$\Phi(G^{(1)}) = A_{T1in} G^{(1)}_{T1in} + A_{T1out} G^{(1)}_{T1out} - A_{T2} G^{(1)}_{T2},$$
$$\Phi(G^{(-1)}) = A_{T1in} G^{(-1)}_{T1in} + A_{T1out} G^{(-1)}_{T1out} - A_{T2} G^{(-1)}_{T2} \qquad (25a)$$

$$\Phi(G^{(1)}) = 0,$$
$$\frac{\partial \Phi(G^{(1)})}{\partial r} = \frac{\sqrt{j(j+1)}}{r} \Phi(G^{(-1)}) \qquad (25b)$$

$\Phi(G^{(1)})=0$ expresses the continuity of the QP TOW displacements at the PNR surface and $\Phi(G^{(-1)})$, the discontinuity of $\Phi(G^{(1)})$ derivatives with respect to ***r*** at the same surface. This discontinuity is due to the excitation of a short-wavelength LO wave, which is important as it leads to the existence of oscillating electric charges in a very thin layer near the PNR surface. Expressions for the functions $G^{(1)}_{T1in}$, $G^{(1)}_{T1out}$, $G^{(1)}_{T2}$, $G^{(-1)}_{T1in}$, $G^{(-1)}_{T1out}$, $G^{(-1)}_{T2}$ can be found in the Appendix. It should also be noted that $G^{(-1)}_{...}$ describe near-field effects and vanishes for large values of the PNR radius. The optic mode displacement, $\boldsymbol{\xi}$, is described by the relation (A1) in the Appendix. The density of electric charge $\boldsymbol{\rho}$=e*div($\boldsymbol{\xi}$) can be described by a superposition of spherical harmonics $\boldsymbol{Y_{jM}(\vartheta,\varphi)}$, ***j***≥1, so that the total charge inside a thin spherical layer is equal to zero, $\oint \rho d\Omega = 0$. Equations (24a, b, 25a, b) allow calculating the phase of scattering, its cross-section and all necessary values (for example local mode frequency, space distribution of displacement in the Vortex or QP TOW excitations).

## 3. RESULTS of CALCULATIONS

### 3.1. *SOFT-MODE - TEMPERATURE DIAGRAM.*

We first plot in figure 1, the soft mode frequency, $\omega(T)$, inside and outside the PNR for the particular values of the transition temperatures, $T_{c2}$ and $T_{c1}$, used in the calculations below. General conclusions about the TOW dynamics are then obtained in the following subsections.

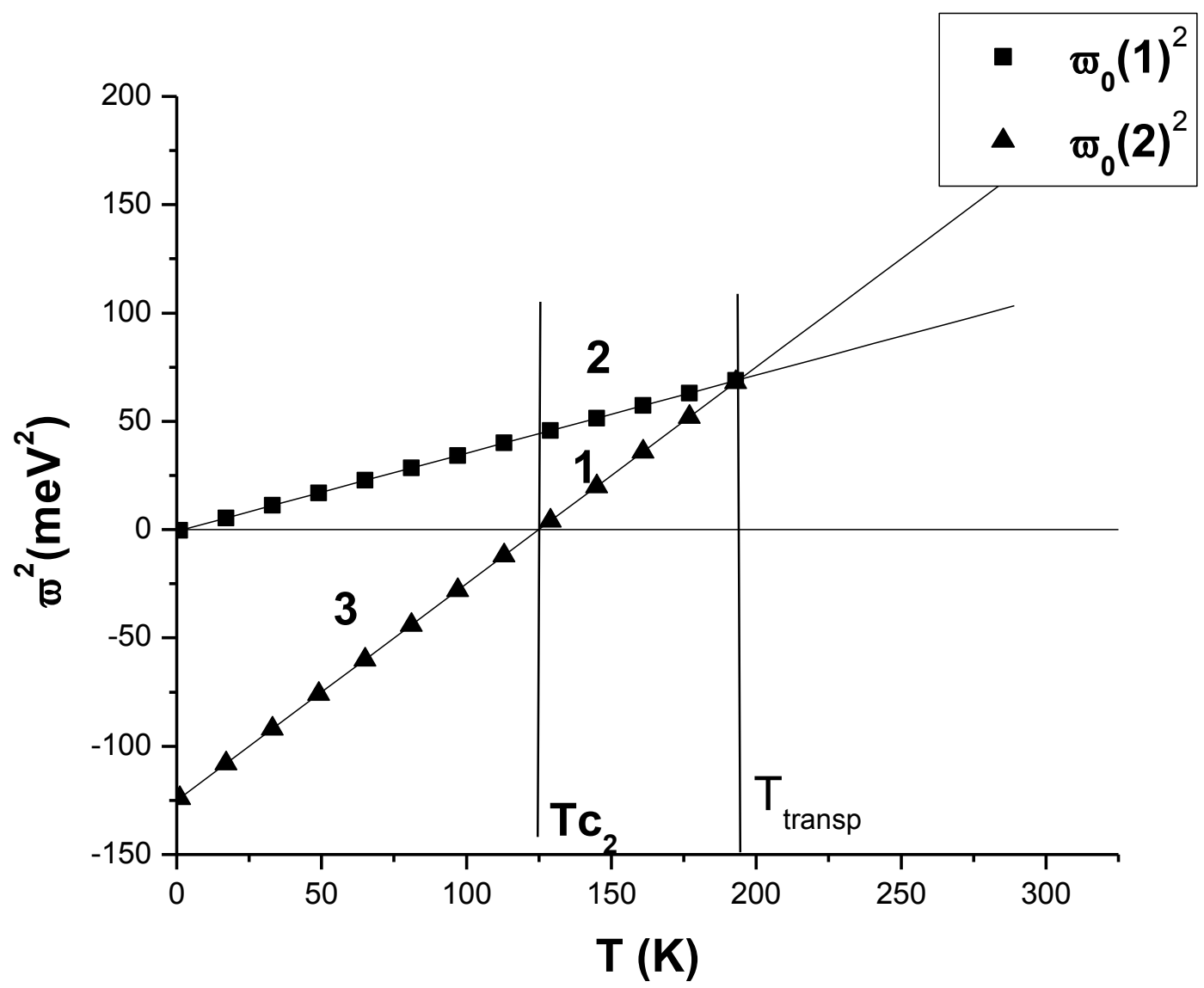

Fig.1 Soft mode energy gap versus temperature for the case of a large radius PNR. $Tc_1$=2K, $Tc_2$=125 K, *a1*=0.6; *a2*=1; $C_O$=70. $\omega_o(1)^2=a1^2(T-Tc_1)$, $\omega_o(2)^2=a2^2(T-Tc_2)$. Numerical values of parameters were estimated for the particular physical systems considered and are being used in following calculations. $T_{transp}$. Is the temperature at which the soft-mode gaps inside and outside the PNR are the same and the soft-mode not scattered by the PNR. $T_{transp}=(a1^2Tc_1-a2^2Tc_2)/(a1^2-a2^2)$, $T_{transp}\approx 194.18$ *K*. 1 – triangle region where localized modes are possible; 2 – region where an incident TOW is scattered by PNR; 3 – condensed PNR phase.

We limited ourselves to the temperature region $T<T_{transp}$.

### 3.2. *SYSTEM INSTABILITY DUE TO VORTEX AND QUASIPOLAR EXCITATIONS*

A PNR embedded in a very large host medium will spontaneously transform to another phase (polar nanodomain, PND) through a local phase transition. A large PNR will undergo a phase transition at $T=T_{c2}$ (Fig.1) but a finite size PNR will transform at a lower temperature $T_{inst}<T_{c2}$. The instability occurs when the frequency ω of the TOW within the PNR goes to zero. In the following, we consider a temperature $T>T_{inst}$.yet transformed to a PND. Therefore, we should find the non-zero amplitude solutions (24a, b) and (25a, b) corresponding to both ω=0 and $A_{T1in}$=0. The boundary conditions are written in the form of a matrix equation: $MA = 0$, where the components of the vector A are the amplitudes $A_{T2}$ and $A_{T1out,}$ respectively inside the PNR and reflected from it outside, and the 2x2 matrix M depends on two parameters – temperature, *T,* and PNR radius, *R*. For example, for the j=1 case, the QP amplitudes are obtained by solving the following coupled equations derived from Eqs.25a and b:

$$A_{T1out}G^{(1)}_{T1out} - A_{T2}G^{(1)}_{T2} = 0 \text{ at } r = R$$

$$A_{T1out}\left(\frac{\partial}{\partial r}G^{(1)}_{T1out} - \frac{\sqrt{2}}{r}G^{(-1)}_{T1out}\right) - A_{T2}\left(\frac{\partial}{\partial r}G^{(1)}_{T2} - \frac{\sqrt{2}}{r}G^{(-1)}_{T2}\right) = 0 \qquad (26)$$

Solving the determinant of Eq.(26) yields several solutions, but only the first ones, corresponding to the highest temperature, are plotted in Fig.2.

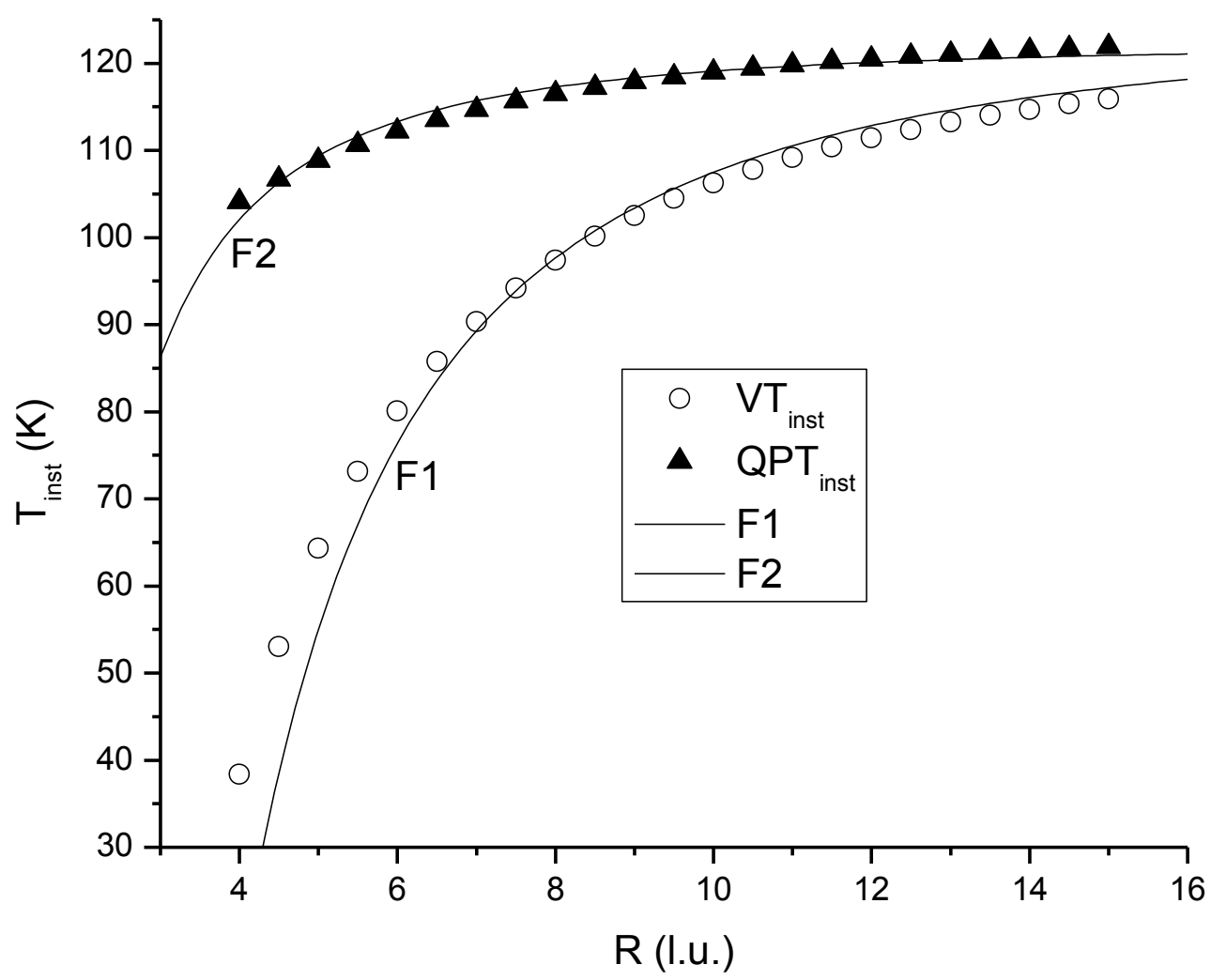

Fig.2. Temperature of the local instability for the case of Vortex, $VT_{inst}$, and Quasi Polar, $QPT_{inst}$ , excitations, respectively, vs PNR radius *R*. Total angular momentum *j*=1. F1, F2 –fitting lines corresponding to y=125-1750/$R^2$ and y =122.34-324.4/$R^2$, respectively.

The temperature of instability is smaller for larger total angular momentum *j*> 1 and for a smaller PNR radius ***R***. For example, for the case of a QP excitation, $T_{inst}$(*j*=2,*R*=4)=73.8K and $T_{inst}$(*j*=2,R=6)=97.5K. This phenomenon is explained by the presence of a "centrifugal barrier", due to the angular momentum contribution ~j(j+1)/$R^2$ in the Hamiltonian. The "average" momentum corresponding to a given density of displacement is larger for the case of a Vortex excitation than for the case of a QP one due to repelling from the PNR center. It is necessary to supercool the PNR below $T_{c2}$ in order to compensate for the corresponding contribution to the free energy. This supercooling is stronger for the case of a Vortex excitation, $T_{inst} \approx T_{c2} - (1 \div 0.25)\frac{C_o^2}{a2*R^2}$ . The respective displacements $\xi_V$ and $\xi_{QP}$ for V and QP excitations can be written in a simple form with spherical coordinates when *j*=1, *M*=0, and at large *r*.

$$\xi_V \sim \frac{\exp(-kr)}{kr}\sin(\theta)\, e_\varphi,\ \ \xi_{QP} \sim \frac{\exp(-kr)}{kr}\sin(\theta)\, e_\theta,\ \ k = a1\sqrt{Tinst - Tc_1}/C_O \qquad (27)$$

A Vortex (***j***=1) excitation describes the rotation of a rigid spherical shell around a polar axis and can be described as a manifold of closed lines (similar to the magnetic field lines in electrodynamics), which does not therefore possess an electric dipole moment. A quasi-polar excitation corresponds to displacements parallel to the polar axis (17b) and integration over spherical angles reveals the existence of a non-zero electric dipole moment but only for *j*=1. Following the previous Eqs (17b) and (26), the dipole moment, ***P***, associated with the QP excitation can be written:

$$P_{\mu} = \chi_{\mu}\{G^{(1)}_{T1out}(R)[G^{(1)}_{T2}(r)\sqrt{2/3} + G^{(-1)}_{T2}(r)\sqrt{1/3}]\theta(R-r) +$$

$$G^{(1)}_{T2}(R)[G^{(1)}_{T1out}(r)\sqrt{2/3}+G^{(-1)}_{T1out}(r)\sqrt{1/3}]\theta(r-R)\},\ \theta(x\le 0)=0,\ \theta(x>0)=1 \quad (28)$$

where G's are defined in Eqs.A6, A7, and A8 in the Appendix. The unit vector $\chi_\mu$ (11) defines the direction of the dipole moment ***P*** and is fixed extrinsically by interaction with chemical short range ordering, dislocations and so on. As shown in Fig.3, the radial dependence of $P_\mu / \chi_\mu$ is mostly confined inside the PNR and sharply decreases at the PNR surface.

In Fig.3, we have calculated the density of TOW displacements for the V and QP excitations near $T_{inst}$. For QP excitations, we calculate separately the densities $DQP_{par}$ for displacements parallel and $DQP_{perp}$ for displacements perpendicular to the radius ***r***. These densities are defined as the square of the displacement averaged over angles inside a spherical layer.

$$DV \equiv \left|f_j(kr)\right|^2,\ DQPpar = \left|G^{(1)}(r)\right|^2,\ DQPperp = \left|G^{(-1)}(r)\right|^2 \quad (29)$$

the PNR surface looks as an almost impenetrable barrier for these displacements $DQP_{par}$ parallel to the radius ***r*** is mostly concentrated inside the PNR such that as compare with the case of vortex excitation.

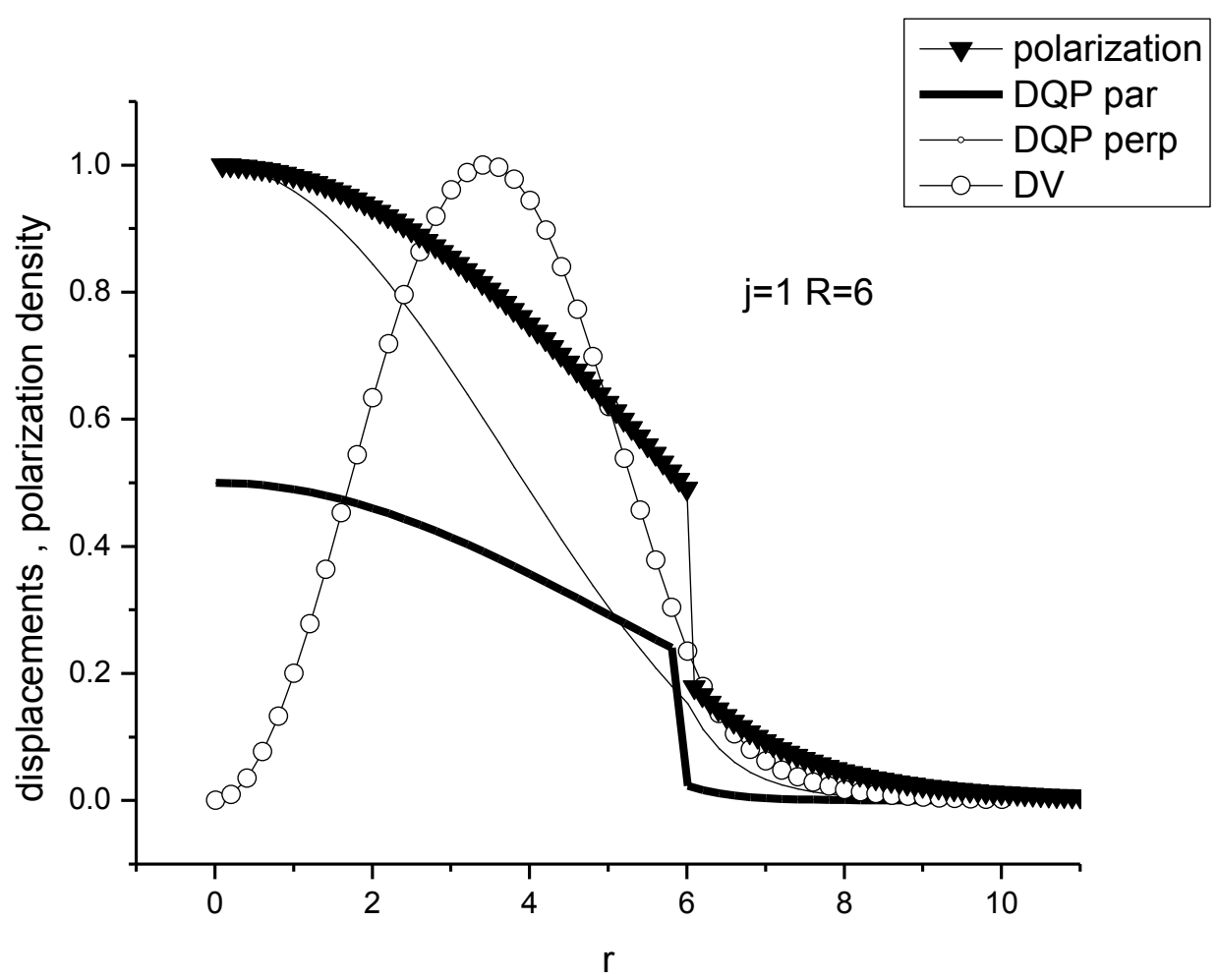

Fig.3 Radial layer distribution density of the normalized QP polarization, QP and Vortex displacements (30) near the temperature of the corresponding local phase transition, $VT_{inst}$=80.1 K, $QPT_{inst}$=112.23K. PNR radius ***R***=6, ***j***=1.

### 3.3 *LOCALIZED AND QUASI LOCALIZED OPTIC MODES*

Local optic modes (LM) are characterized by an infinite lifetime as determined by Im(ω)=0. They are in general discrete and exist in the "cage" created by the host media; the LM frequency is lower than the frequency of modes propagating outside the PNR, $\omega_{LM} < \omega_0(1)$. Therefore, LMs cannot be "directly" excited by a propagating incident TO wave and only shallow LMs can scatter it effectively. Quasi localized modes, QLM, have an energy that is greater than the soft mode energy gap of the host medium. They result from the effective long-time interaction between an incident TOW and the PNR. They are seen as higher amplitude of the propagating TO mode in the vicinity of the PNR and can lead to its resonant scattering. The analysis of LM and QLM modes is presented successively in the following sections.

*Local optic modes*

Vortex LMs appear if a non-zero amplitude solution of Eqs (24a), (24b) exists when $A_{in}^{(0)}=0$. Similarly, QP LMs appear if a non-zero amplitude solution of Eqs (25a), (25b) exists when $A_{T1in}=0$. These become respectively:

$$\frac{1}{g_j}\frac{\partial g_j}{\partial r}=\frac{1}{g_{j,out}}\frac{\partial g_{j,out}}{\partial r},\ r=R,\ Im(\omega_{LM})=0,\quad Im(q1)>0 \qquad (30a)$$

$$\frac{1}{G_{T2}^{(1)}}\frac{\partial G_{T2}^{(1)}}{\partial r}-\frac{\sqrt{j(j+1)}}{r}\frac{G_{T2}^{(-1)}}{G_{T2}^{(1)}}=\frac{1}{G_{T1out}^{(1)}}\frac{\partial G_{T1out}^{(1)}}{\partial r}-\frac{\sqrt{j(j+1)}}{r}\frac{G_{T1out}^{(-1)}}{G_{T1out}^{(1)}} \qquad (30b)$$

Two solutions of these, high and low frequency, are shown are shown in Figs 4a

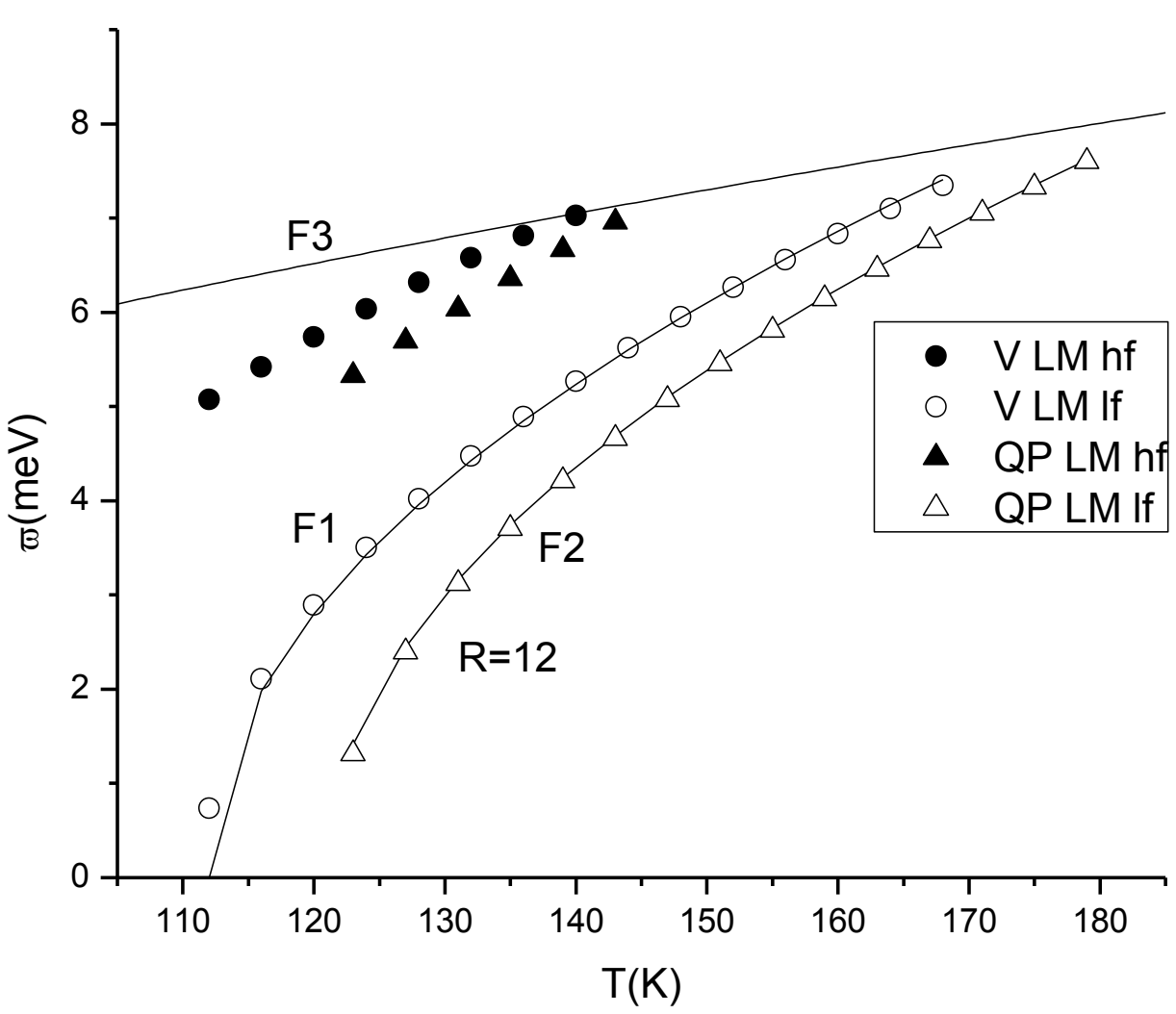

Fig.4a. Frequency ω of the Vortex and QP localized excitations, LM. The high (hf) and low (lf) frequency branches are shown as a function of temperature. The PNR radius was chosen to be *R*=12 and *j*=1. The QP mode becomes unstable at T≈121K. Curves F1 and F2 are fitting curves corresponding to $y=(T-112)^{1/2}$ and $y=(T-121)^{1/2}$ respectively. Also shown is the temperature dependence of the host medium soft mode gap, F3, which sets an upper limit for the value of the LM frequency.

High and low frequency LMs correspond to different radial distributions of the displacement for a given value of the total angular momentum ***j*** (see later) and these frequencies increase with ***j***. They are analogous to the different discrete levels corresponding to a given radial quantum number in quantum mechanics [17]. The temperature dependence of the low frequency LM branch, $\boldsymbol{\omega_{LMlf}}$ , can be explained as follows. The typical momentum inside the PNR is proportional to b/R (b being a numerical constant) and the LM excitation does not penetrate deeply into the host medium. Taking into account expressions (9) and (5), we can write the frequency of the LM as : $\omega_{LMlf} \approx \sqrt{(\frac{b}{R})^2 + (a2)^2(T - Tc_2)}$ or equivalently $\omega_{LMlf} \approx a2\sqrt{T - Tc_2^*}$, with $Tc_2^* = Tc_2 - (\frac{b}{a2*R})^2$. The temperature dependence of the low frequency LM branch is thus similar to that of the soft-mode gap (*q*=0) but with a smaller value of $T_{c}$. The frequency of the local QP mode is smaller than that of the Vortex for two reasons: 1) the QP wave field (*j*=1) contains a component with zero orbital momentum (15), and 2) QP mode interacts with the high energy LOW at the PNR surface. This interaction leads to a decrease of the QP mode energy, in accordance with the general results of second order perturbation theory. [17] Although not shown here, low and high frequency Vortex local modes correspond to different spatial distributions of the TOW displacement. Sample calculations for the following parameters (*R*=12, *j*=1, *T*=140 K, $\omega_{LMlf}$=5.268, $\omega_{LMhf}$ = 7.028, $\omega_0(1)$=7.048, $\omega_0(2)$ = 3.87 meV) reveal a maximum of the low frequency LM displacement density at r≈6.6 and only one node at the PNR center. The high frequency Vortex LM displacement density has 2 maxima at r≈4 and r≈11.2 and an additional density node at r=8.61 inside the PNR and the local h.f. Vortex mode is shallow in this case. This picture – increasing number of nodes with increasing energy - is similar to the dependence of wavefunctions on the radial quantum number in atomic physics [17].

The QP local mode structure is shown in Fig. 4b.

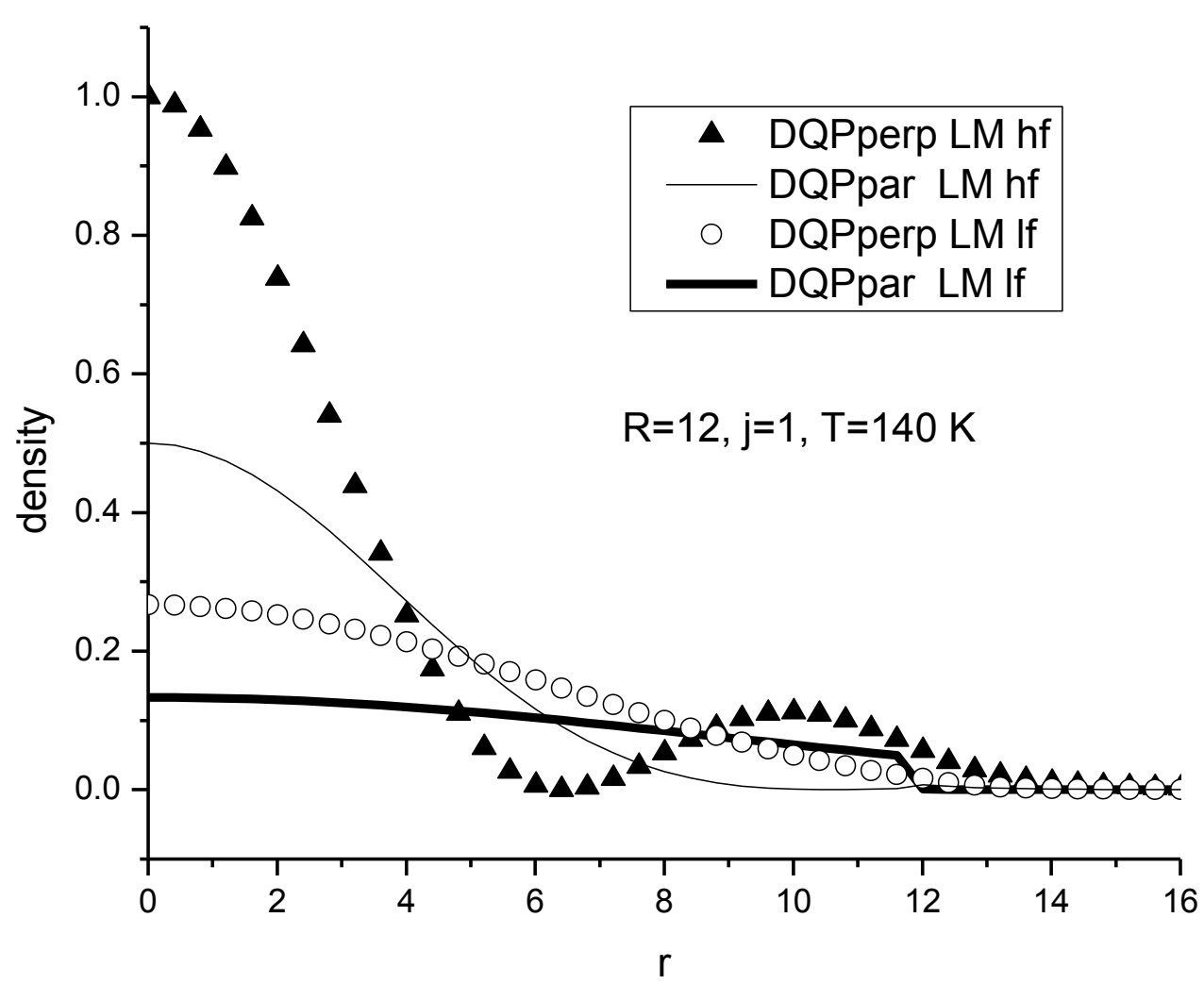

Fig.4b. Radial layer distribution density of the normalized QP displacements for the case of h.f. and l.f local modes. *R*=12, *j*=1, *T*=140 K, QP $\omega_{LMhf}$=6.123, $\omega_{LMlf}$=4.384 meV.

Both DQPperp and DQPpar have a maximum at the PNR center. The displacement density for DQPperp of the h.f. mode has an additional node at r=6.41. The most unusual behavior is that of the h.f. and l.f. LM displacements parallel to ***r*** **:** both show no node and the density discontinuity at the PNR surface is stronger for the low frequency LMs. LMs exist within a limited temperature interval. For example, in the case of smaller PNR size *R*=6, Vortex LMs only exist for T≤140K. The structure of the TOW localized in PNRs could be studied by means of inelastic neutron scattering.

*Quasi local optic modes*

It is impossible to directly "convert" a single propagating incident TOW to a local mode because the frequency of the incident wave is more than that of the LM frequency. However a TOW can be resonantly scattered by the PNR due to the existence of the so-called quasi local modes, QLM, the frequency of which appears in the continuous TOW energy spectrum. These modes are due to the long-time effective interaction between incident TOW and PNR, and lead to resonant TOW scattering. QLMs are similar to quasi-stationary states in quantum mechanics [17] and correspond to the solution of the dynamical equation, taking into account only the emitted (outgoing) waves $\propto \exp(iq_1 r - i\omega t)$. This solution is *divergent* at large distances from the center of the PNR, ($Im(q_1)<0$), and its frequency ω contains a small *negative* imaginary part which corresponds to the QLM decay. QLM's are described by the solutions (24a,b), (25a,b) corresponding to Vortex and QP excitations respectively, jointly with the condition $Im(\omega)<0$, $Im(q_1)<0$. The QLM momentum, $\boldsymbol{q_1}$, and frequency are shown in Table 1 and Fig 5.

Table 1. Parameters of a few Vortex and QP local (LM) and quasi local modes (QLM) for *R*=6 and *j*=1.

| Excitation | T,K | Re(ω), meV | -Im(ω), meV | Re(q1), r.l.u. | -Im(q1) ), r.l.u. | Re(q2) , r.l.u. | -Im(q2) , r.l.u. |
|---|---|---|---|---|---|---|---|
| V LM | 130 | 6.51 | 0 | 0 | -0.027 | 0.08734 | 0 |
| V QLM | 150 | 7.42 | 0.668e-1 | 0.190e-1 | 0.715e-3 | 0.783e-1 | 0.174e-3 |
| V QLM | 170 | 8.01 | 0.387 | 0.275e-1 | 0.286e-2 | 0.625e-1 | 0.126e-2 |
| QP QLM | 130 | 6.954 | 0.026 | 0.0239 | 0.0107 | 0.09407 | 0.0027 |
| QP QLM | 150 | 7.44 | 0.0269 | 0.024 | 0.0126 | 0.07875 | 0.00385 |
| QP QLM | 170 | 7.856 | 0.0285 | 0.02236 | 0.016 | 0.05864 | 0.0061 |

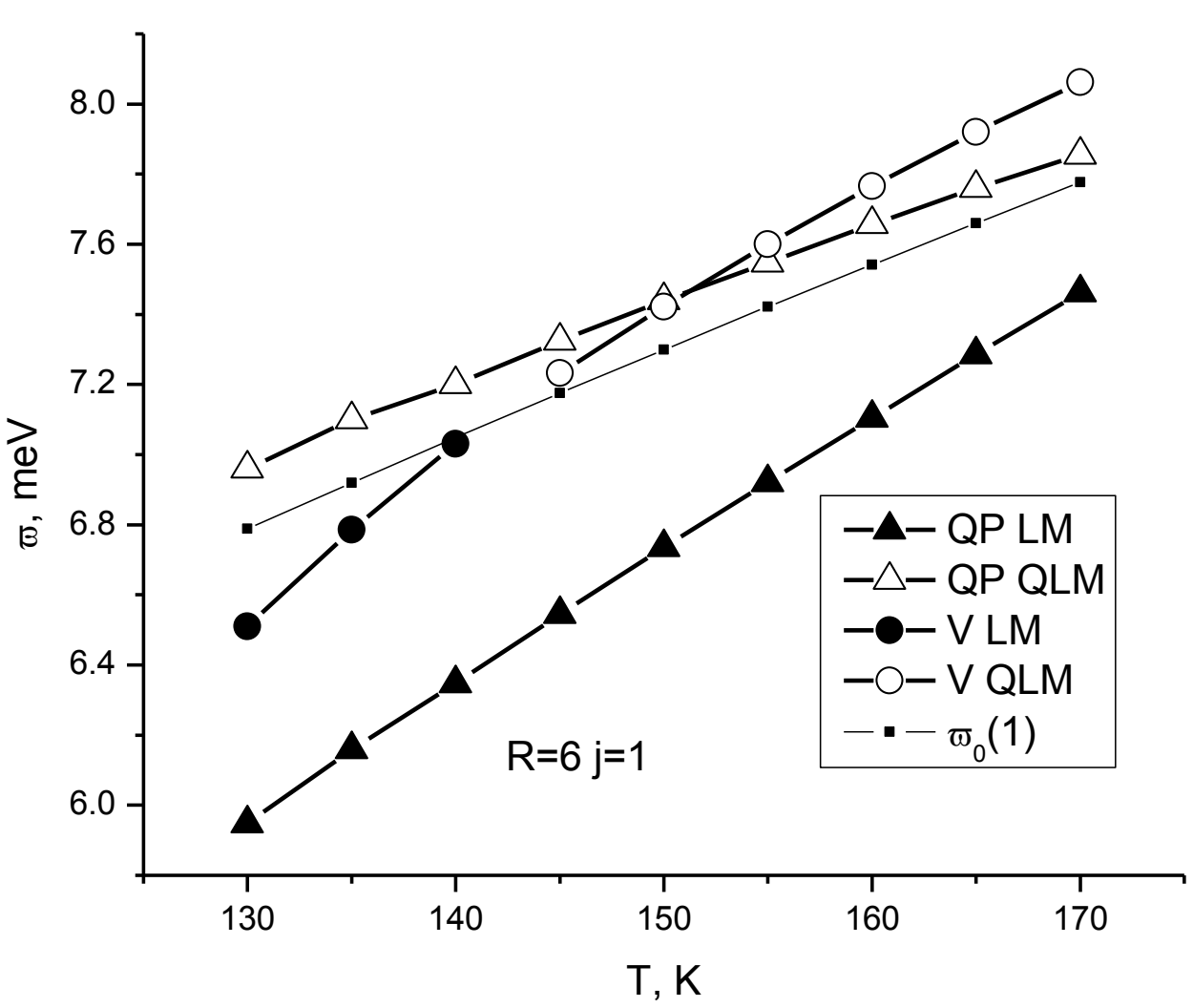

Fig.5. QP and Vortex local, LM, quasi local mode, QLM, frequency, and host medium soft mode gap $\boldsymbol{\omega_0(1)}$ as a function of temperature. *R*=6, *j*=1.

Localized Vortex modes (V LM) exist for T≤140 K and "convert" to QLMs for T>140K (Fig.5). Quasi polar LM and QLMs co-exist in the range 130K ≤ *T*≤ 170K.

Because the value of |Im(q2)/Re(q2)|is small, Vortex and QP QLMs correspond to "free" TOW motion inside the PNR. The parameters characterizing them strongly depend on temperature.

3.4. *TRANSVERSE OPTIC WAVE SCATTERING BY THE POLARIZED NANO REGION.*

The scattering cross-sections of transverse QP and V excitations have been calculated using Eq.(22). Figs. 6- 9 illustrate their dependence on wavevector and temperature.

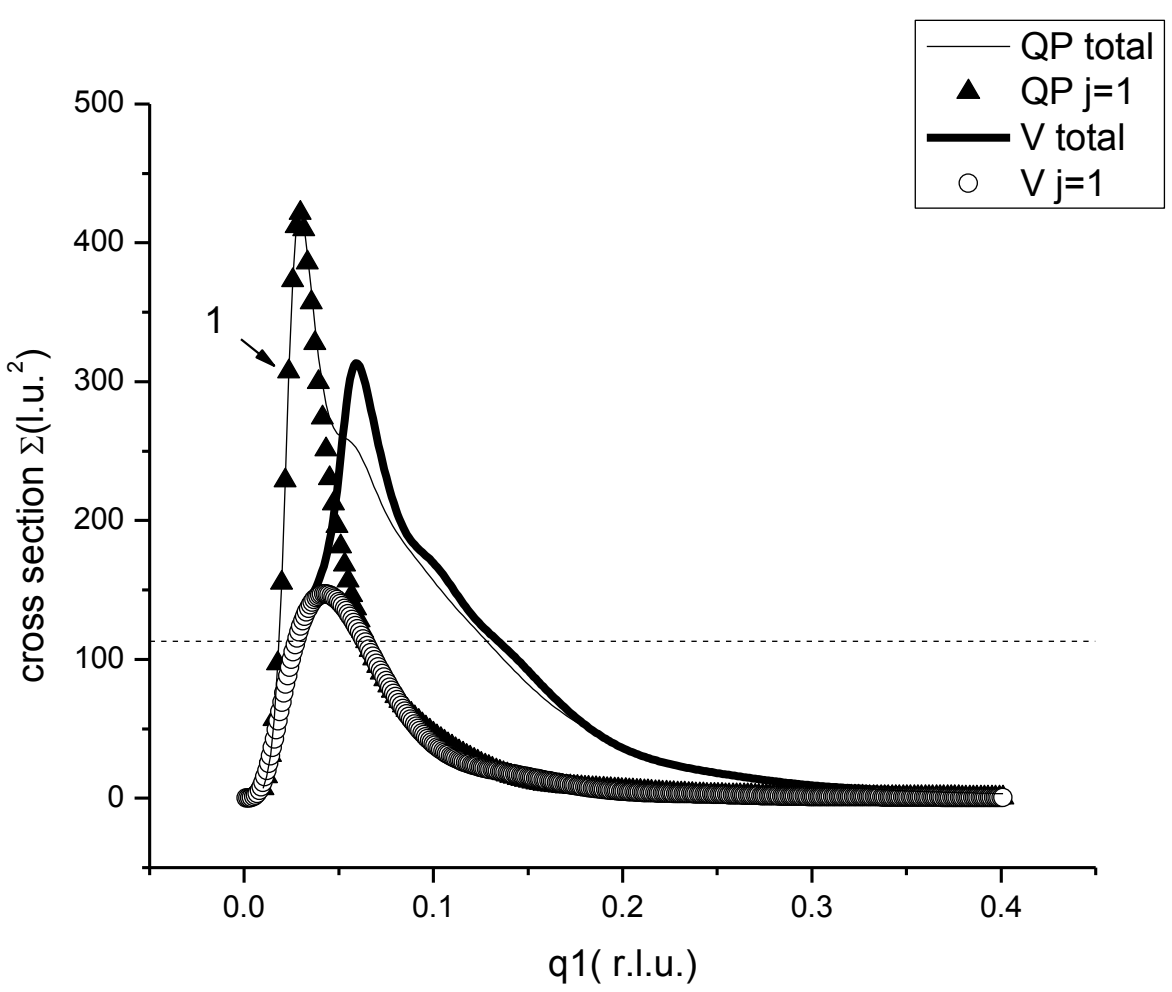

Fig.6. QP and Vortex scattering by PNR (total and partial, *j*=1, cross section Σ) showing a strong resonance. The incident TOW momentum $\boldsymbol{q_1}$ is expressed in units $\pi/\boldsymbol{a}$ where $\boldsymbol{a}$ is a value similar to the lattice constant and *R*=6, *T*=130 K, $C_O$=70 meV/r.l.u.; 1- marks the position of the QP quasi-localized state; the horizontal dash line marks the PNR geometrical cross-section.

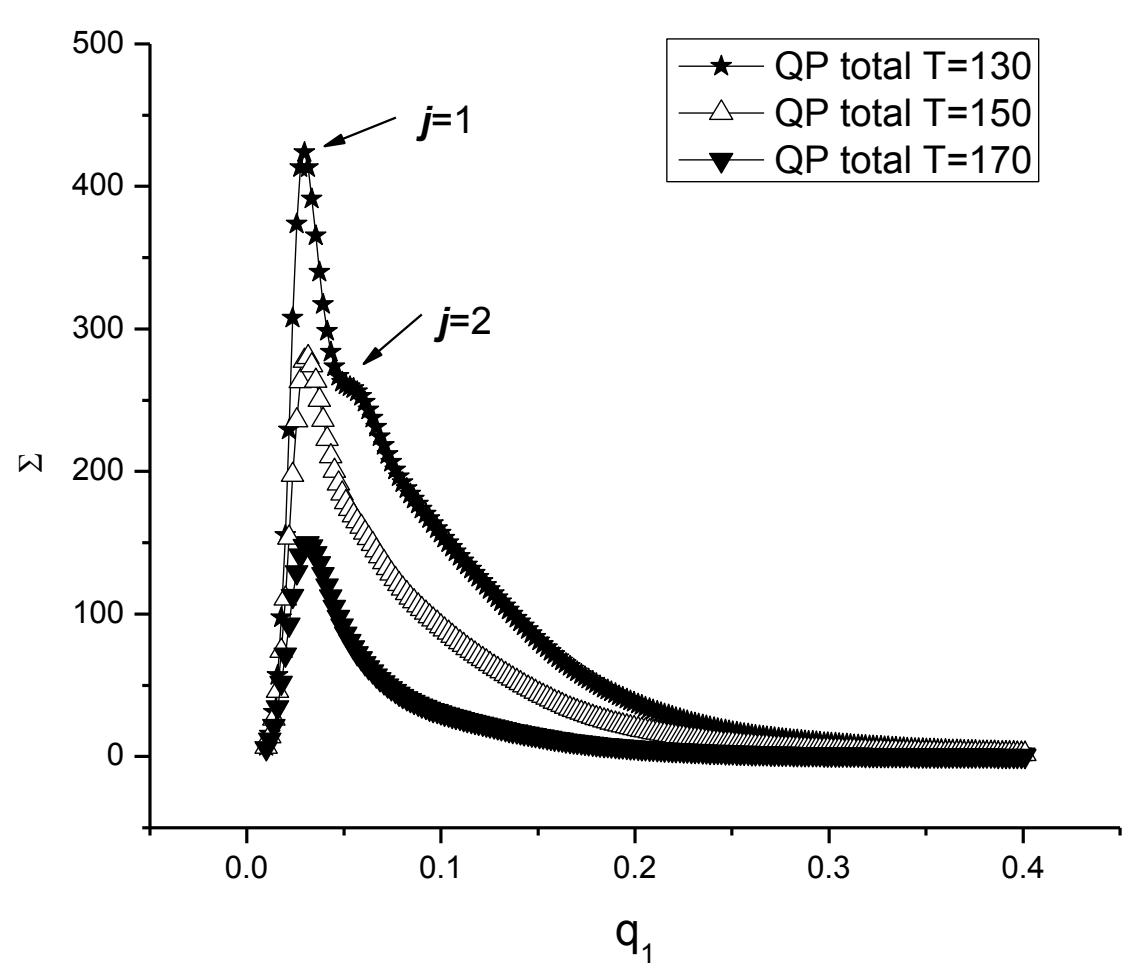

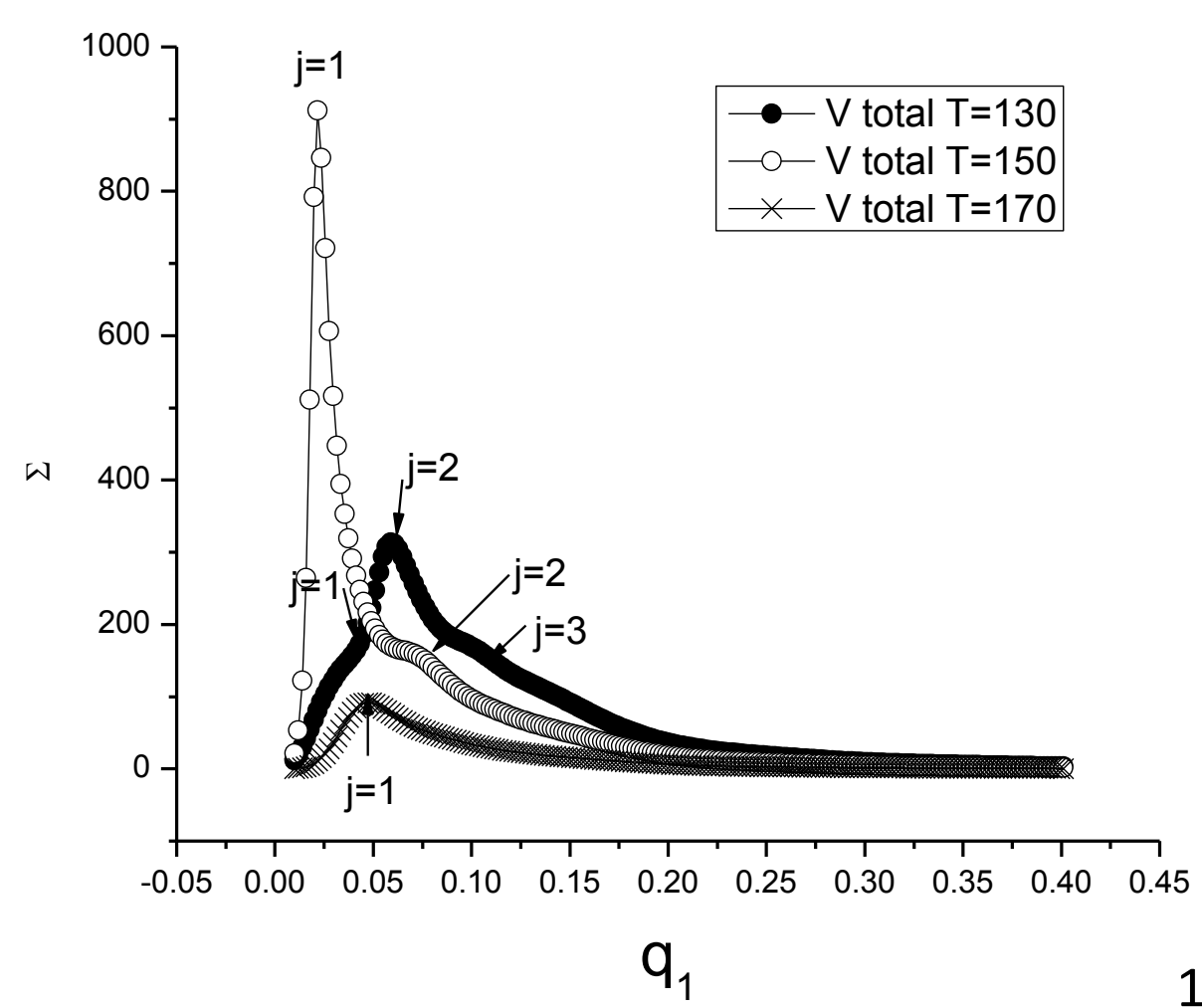

a) b)

Fig.7 a). Total cross section of PNR as function of the incident momentum $\boldsymbol{q_1}$ at the temperature T=130, 150, and 170 K. a) for scattering by QP mode, long tail part of curves ($\boldsymbol{q}_1$ >0.15) can be approximated by Σ(T=130K)=0.15/$\boldsymbol{q}_1^{3.5}$ , Σ(T=150K)=0.065 /$\boldsymbol{q}_1^{3.6}$ , Σ(T=170K)=0.022/$\boldsymbol{q}_1^{3.6}$ ; b) for scattering by for scattering by Vortex mode, the long tail part of curves ($\boldsymbol{q}_1$ >0.15) can be approximated by Σ(T=130K)=0.15/$\boldsymbol{q}_1^{3.5}$ , Σ(T=150K)=0.075 /$\boldsymbol{q}_1^{3.5}$ , Σ(T=170K)=0.02/$\boldsymbol{q}_1^{3.5}$

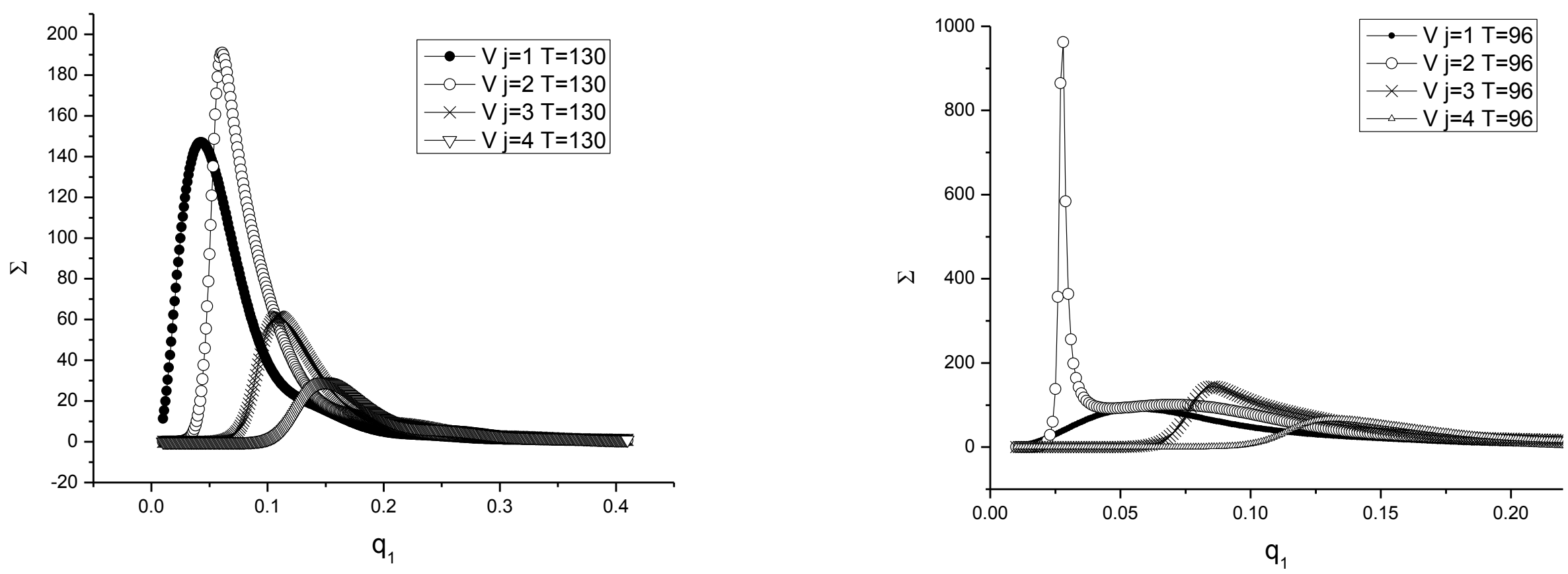

a) b)

Fig.8. Vortex partial-wave scattering cross sections, corresponding to $\boldsymbol{j}$=1, 2, 3, 4 and R=6 a) at T=130 K and b) 96K, We note in particular the dominant role of the quadruple ($\boldsymbol{j}$=2) scattering.

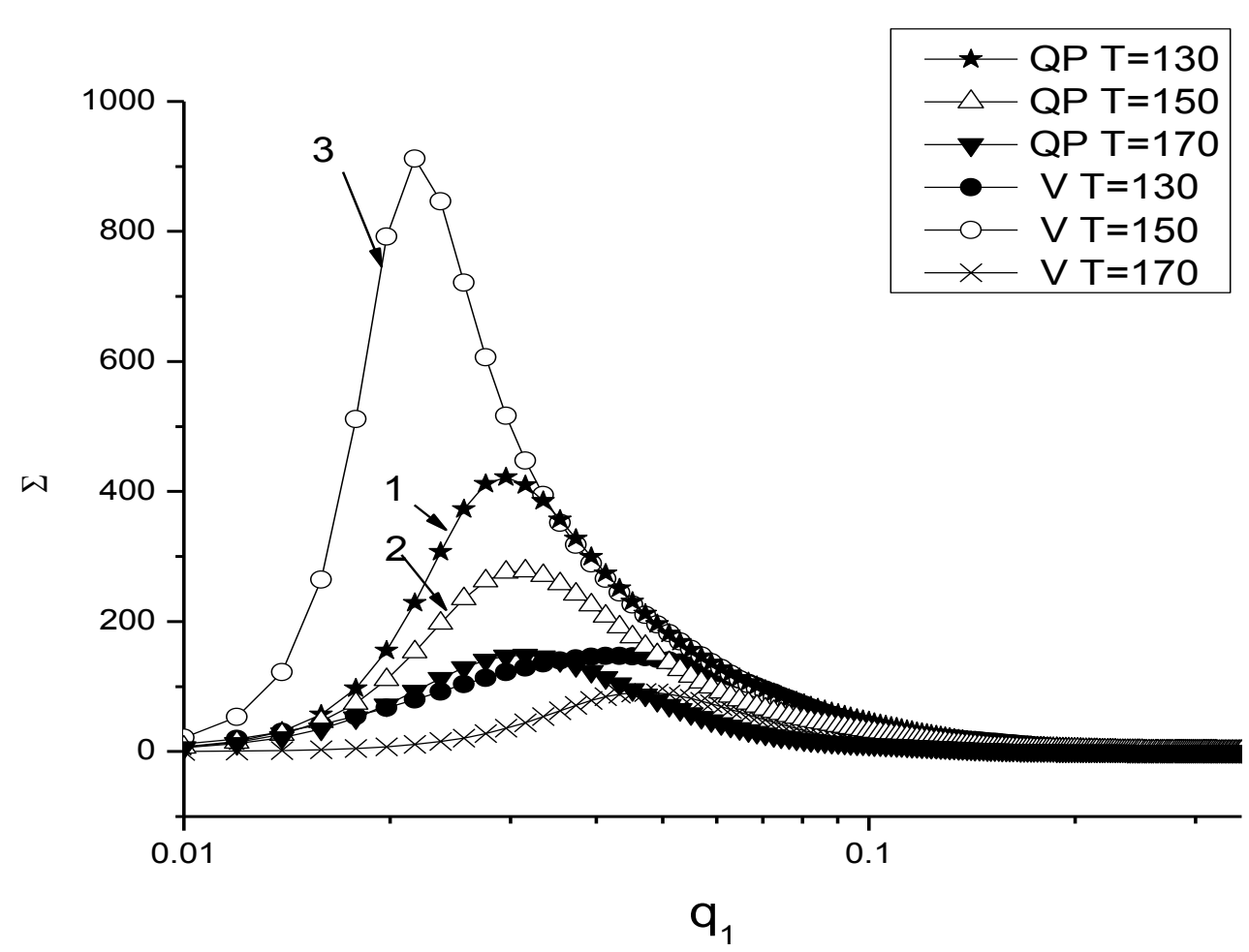

Fig.9. Partial (***j***=1) QP and Vortex scattering cross section as function of $q_1$ with angular momentum at the temperature 130, 150, and 170 K. ***R***=6, $C_O$=70 meV/r.l.u.; 1,2, and 3 – mark the positions of the QP and Vortex quasi-localized states.

The TOW scattering obviously also depends on the PNR radius (Fig.10).

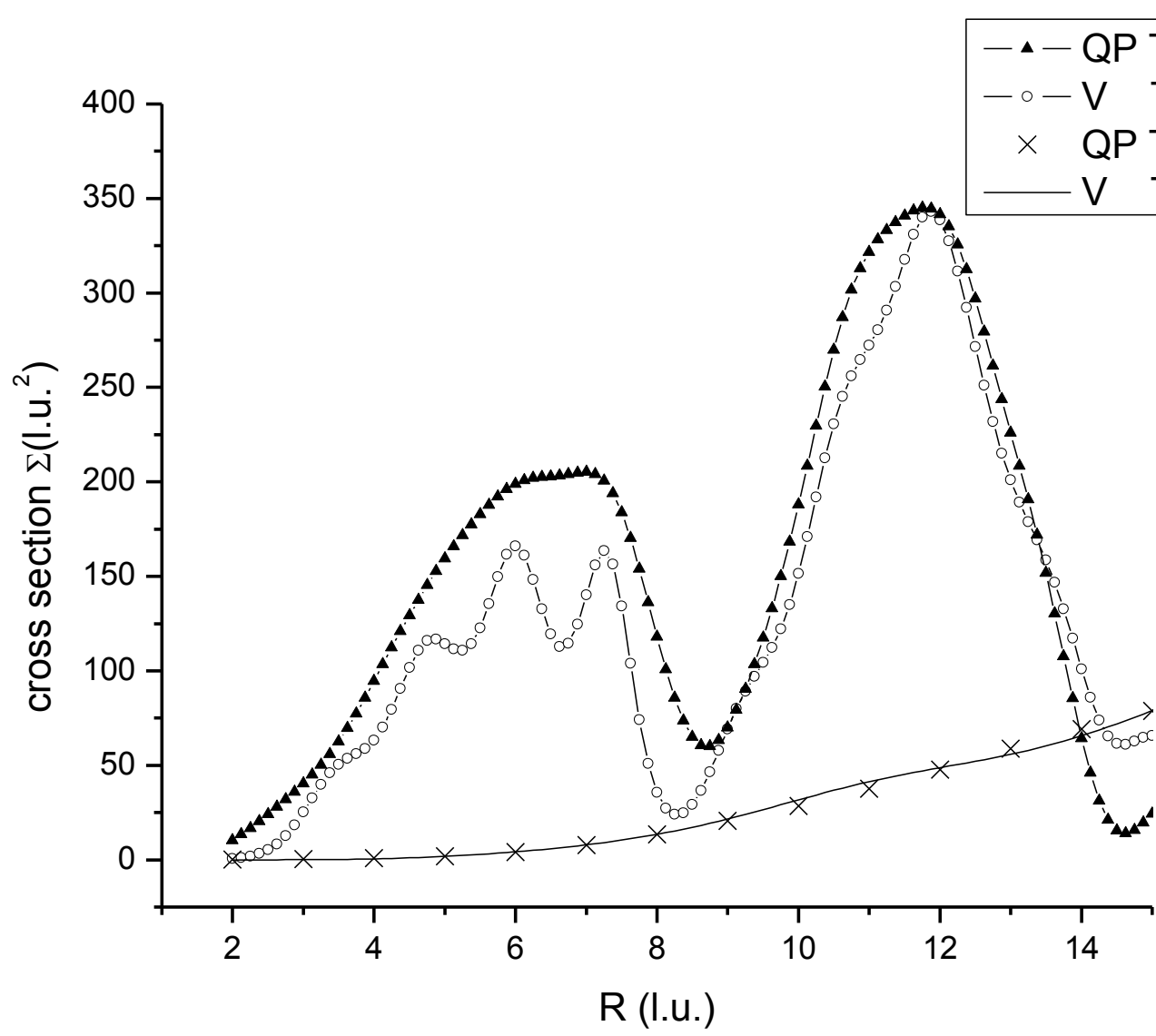

Fig.10. QP and Vortex total scattering cross section as a function of PNR radius value ***R*** : ***q1***=0.1; ***T***=130 and 190K.

Diffraction maxima and minima are seen here for the both V and QP scattering at T=130K. Such maxima and minima are absent at T=190K because the soft-mode gaps inside and outside the PNR have almost of the same value at higher temperature (see Fig.1) and the scattering there is weak. The moderate oscillations in the curve corresponding to Vortex scattering at T=130K are created by partial harmonic contribution to the cross-section.

## 3.5. DISCUSSION

The calculated results shown in Figs. 6 - 9 can be interpreted qualitatively as follows.
For the case of very small incident momentum $q_1$, the TOW dynamics is similar to the scattering of a slow particle in quantum mechanics [17]. The calculated value of the scattering cross-section is very small, $\sigma \sim q_1^4$ for $q_1 \leq 0.01$ and $j$=1, which is in agreement with the general theory of scattering [17].
The scattering cross section Σ decreases at large momentum, $q_1$>0.15 (Fig. 7a, b). For the case of QP and Vortex excitation Σ(T=130K) ≈ $0.15/q_1^{3.5}$ , Σ(T=150K) ≈ $0.065/q_1^{3.6}$ , Σ(T=170K)= ) ≈ $0.022/q_1^{3.6}$ and Σ(T=130K) ) ≈ $0.15/q_1^{3.5}$ , Σ(T=150K) ) ≈ $0.065/q_1^{3.6}$ , Σ(T=170K) ) ≈ $0.022/q_1^{3.6}$ , respectively.
Therefore the value of the potential scattering cross section has a wide maximum in the intermediate region around $q_1 \sim 0.05 \div 0.1$. The reason for the decrease of the cross section at large momentum is as follows. Scattering is due to the difference in the soft mode gaps (Eq.5) inside and outside the PNR, $\Delta\omega \equiv \omega_0(1) - \omega_0(2)$, which is equal to 4.57, 2.3 and 1.08 meV at 130K, 150K and 170 K respectively, and defined by the value of the dimensionless parameter $\Delta\omega/C_0 q_1$ <1. This potential scattering is stronger at low temperature and decreasing for large values of $q_1$. The effect of scattering disappears at the temperature $T_{transp}$. (see Fig1) at which the soft-mode gaps inside and outside the PNR become the same and the soft-mode is no longer scattered by the PNR, $T_{transp} \approx 194$ *K* .Then, QP and V scattering is small and almost the same at ***T***=190K (Fig.10).

However this picture of the potential scattering does not exhaust the problem. As shown in Figs. 6-9, strong resonant scattering of the TOW is clearly seen at small q. There, the value of the scattering cross section is 4 ÷ 8 times larger than the value of the geometrical PNR cross section $\pi \boldsymbol{R}^2 \approx 113$ l.u.$^2$.

The difference between QP and Vortex scattering is essential in the resonance region, due to the effective long time interaction between excitation and PNR. QP and V cross sections are similar outside of this region where the corresponding cross section can be calculated in the Born approximation and with a plane wave basis.

The calculated *total* scattering cross section as function of the momentum $\boldsymbol{q_1}$ exhibits a strong peak asymmetry and bumps. These peculiarities are the results of scattering QP and V excitations with different angular momentum ***j*** (see especially Fig.8). These bumps are more visible at low temperature.

The relative value of the partial cross section depends on temperature. For example, Vortex peaks at are smaller than QP ones T=130 K and 170 K but not at T=150K case (compare Figs.7a and b). This is due to the presence of quasi local (QLM) and local (LM) modes. As explained above, the propagating incident TOW can "directly" excite QLMs, leading to strong resonance scattering. Similar resonant scattering is not possible by LMs because their energy is smaller than the energy of the incident TOW. In general, only shallow LM will have a significant effect on scattering. The difference in the effect of LMs and QLMs on scattering is confirmed by a comparison of scattering corresponding to ***j***=1 (Fig.9) and data shown in Fig. 5 calculated for the same case ***j***=1. QP LMs are not shallow and probably do not result in a strong scattering effect. The q1 position of the QP quasi local modes position is almost independent of temperature and is found near the scattering peak at T=130 and 150 K (Fig. 9). By contrast, the position of the Vortex QLM is found at small q1, Re(q1)≈0.019, and near the strong scattering peak maximum at T=150 K, but it moves away at higher temperature (Re(q1)≈0.0275 at T=170 K) when the role of resonance scattering and the value of the cross section correspondingly decrease. Only shallow Vortex LMs exist at low temperature, T=130K, leading to only moderate V scattering.

Finally, we should note that the position of the V cross section maximum along the q1 axis depends on temperature unlike that of the QP cross section (Fig.9). This fact is confirmed by the position of the QP and V local and quasi local modes in Fig.5.

In Figs.6 and 9, the fact that the QLM position and maximum of the scattering cross section do not coincide is due to interference between resonance and potential scattering. In Fig.8a, the partial cross section peaks corresponding to angular momentum ***j***>1 appear at higher $q_1$ positions because the centrifugal force repels the TOW from the PNR. In practice, resonance scattering is most important for *j*=1 and ***j***=2. The appearance of QLM for ***j***=2 leads to the dominant role of the Vortex quadrupole scattering at ***T***=96K (Fig.8b).

## 4. CONCLUSION

We have developed a model describing the dynamics of the Transverse Optic Wave (TOW) in a medium containing one spherically symmetric Polarized Nano Region (PNR). Our analysis is limited here to the temperature region below Burns temperature but above the temperature when PNRs become static or quasi-static polarized nano domains (PND). Two separate types of excitations are found, Vortex and Quasi-Polar (QP). Vortex excitations correspond to closed polarization lines while QP excitations correspond to open polarization lines with a dipole moment (if the total angular momentum *j*=1). Standard boundary conditions are applied at the PNR surface in the case of Vortex excitations. For quasi polar excitations, after excluding the Longitudinal Optic Wave (LOW) contribution, the boundary conditions correspond to continuous TOW displacements but discontinuous derivatives. The discontinuity is connected with near field effects at the PNR surface where the transverse optic mode mixes with the longitudinal one.
We show that a local violation of stability ("phase transition in PNR") occurs at a temperature $T_{inst}$, the value of which is smaller in the case of Vortex than QP excitations. Therefore, the QP condensation occurs at a higher temperature than the Vortex condensation upon cooling. We also find that a localized TO mode (LM) can exist inside PNR and calculate its LM frequency, $\boldsymbol{\omega_{LM}}$. A local phase transition then takes place when $\boldsymbol{\omega_{LM}}\approx 0$. We have computed the cross-section for scattering of a TOW by a PNR and found a strong resonance. Vortex and QP scatterings are practically the same outside the resonance region. In the resonance region however, the Vortex scattering is significantly different from the QP one, which is attributed to the presence of localized and especially quasi-localized modes. The structure of these modes could be studied by means of inelastic neutron scattering.

A detailed discussion of the experimental situation is out of scope for our paper. We note that Gehring et al [6, 9] found evidence of a strong step-like increase in TOW damping (waterfall) for an incident TOW momentum $\boldsymbol{q_1<q_0}$. The value of $\boldsymbol{1/q_0}$ was found to be consistent with the size of PNRs. We also find a broad TOW scattering peak around momentum $\boldsymbol{q_0}$. In addition however, our calculations reveal the presence of a strong increase in TOW scattering at small $\boldsymbol{q_1}$, due to the presence of shallow localized and especially quasi localized states within and around PNRs. The value of the calculated resonance scattering cross section is 4 ÷ 8 times greater than the PNR geometrical cross section. At small $\boldsymbol{q_1}$ the model of scattering by individual and well separated PNRs is no longer valid since interference between waves scattered by different PNRs becomes essential. Our calculation shows that strong soft mode scattering is present in the paraelectric phase. Probably, in order to have a waterfall, it is not necessary to assume that PNRs are in a condensed, ferroelectric phase (PND); it is sufficient that the system contain micro regions, each of size ~ 10 ÷ 20 lattice constants, with values of the soft-mode energy gap different from that of the host medium. This, of course, does not exclude the important role of PNDs in TOW scattering at lower temperatures. However, our calculations are limited here to the temperature range $\boldsymbol{T}>T_{inst}$. An analysis of the "condensed phase" ($\boldsymbol{T}<T_{inst}$) is more complicated since anharmonic interactions will be important in that case.

## APPENDIX.

### DEDUCTION OF BOUNDARY CONDITIONS FOR THE CASE OF QUASI POLAR EXCITATIONS

First of all we discuss boundary condition terms (7b), (7c) including the electrostatic potential $\boldsymbol{\varphi}$. We have solution of Eqn. (7b) outside of PNR surface taking into account Eq-s (19, 20a, 20b)

$$\Delta\varphi - 4\pi e^{*}(\nabla_{\alpha}\xi_{\alpha}) = 0,\ \xi_L = \nabla\Phi_0,\ \Phi_0 = G_j(r)Y_{jM}(\vartheta,\phi),$$

$$\varphi=4\pi e^{*}\Phi_0+\delta\varphi,\ \Delta\delta\varphi=0,\ \delta\varphi\equiv\delta G_j(r)Y_{jM}(\vartheta,\phi),$$

$$\frac{\partial}{\partial r}\left(r^2\frac{\partial\delta G_j(r)}{\partial r}\right) - j(j+1)\delta G_j(r) = 0 \qquad \text{(A1a)}$$

Solutions of (A1a) which is limited inside and outside PNR are well known

$$\delta G_j(r)=A_{\delta}^{(2)}r^{j}\theta(R-r)+A_{\delta,-1}^{(1)}r^{-j-1}\theta(r-R),\theta(x<0)=0,\ \theta(x>0)=1 \qquad \text{(A1b)}$$

where $\boldsymbol{A_{\delta}^{(2)}}$, $\boldsymbol{A_{\delta,-1}^{(1)}}$ –arbitrary constants and $\boldsymbol{R}$- PNR radius.

The boundary condition parts of (7b), (7c) lead to Eqs (A1c), (A1d)

$$A_{\delta}^{(2)} jR^{j-1} - 4\pi e^{*} f_{j,-1}^{(2)} = -A_{\delta,-1}^{(1)} (j+1) R^{-j-2} - 4\pi e^{*} f_{j,in,-1}^{(1)} - 4\pi e^{*} f_{j,out,-1}^{(1)} \quad \text{(A1c)}$$

$$4\pi e^{*} G_{j}^{(2)}(R) + A_{\delta}^{(2)} R^{j} = 4\pi e^{*} G_{j}^{(1)}(R) + A_{\delta,-1}^{(1)} R^{-j-1} \quad \text{(A1d)}$$

where $\boldsymbol{f}^{(2)}_{j,-1}$, $\boldsymbol{f}^{(1)}_{j,in,-1}$, $\boldsymbol{f}^{(1)}_{j,out,-1}$ – transverse optic wave displacement component along the PNR radius. It is interesting to note that *longitudinal* optic wave amplitudes are absent in (A1c). We can solve (A1c, A1d) with respect to $A_{\delta}^{(2)}$ and $A_{\delta,-1}^{(1)}$ and satisfiy the BC's (7c) and (7d). The electrostatic potential $\boldsymbol{\varphi}$ is included in Eqn. (7a) in the form $\boldsymbol{\partial\varphi/\partial r_\alpha}$ and therefore does no affect on the circulation ***curl***($\xi$) dynamics at the TOW. Due to the equality $\Delta\boldsymbol{\delta\varphi}=0$, the electrostatic potential component, $\boldsymbol{\delta\varphi}$, does no affect on the electric charge *e*div(ξ)* dynamics at the LOW. Therefore we omit boundary conditions (7b), (7c) for the electrostatic potential in the reasonably approximation.

In order to satisfy other boundary conditions, it is necessary to take into consideration functions (17a, b, c), (19). These functions contain two components, proportional to the function $\boldsymbol{Y}^{(1)}_{jM}$ and $Y^{(-1)}_{jM}$ respectively. The contribution to the optic wave displacement of the corresponding components can be written as follows:

$$\boldsymbol{\xi} = G^{(1)}(r)\mathbf{Y}_{jM}^{(1)} + G^{(-1)}(r)\mathbf{Y}_{jM}^{(-1)} \quad (A2)$$

The "force" terms of the boundary conditions (7a) contain expressions (A3) and (A4).

$$div(\xi) n_{\alpha} = Y_{jM,\alpha}^{(-1)} [-\frac{\sqrt{j(j+1)}}{r} G^{(1)} + \frac{\partial}{\partial r} G^{(-1)} + \frac{2}{r} G^{(-1)}] \quad \text{(A3)}$$

$$\tilde{\xi}_{\alpha\beta} n_{\beta} = Y_{jM,\alpha}^{(1)} / 2 * [\frac{\partial G^{(1)}}{\partial r} + \frac{\sqrt{j(j+1)}}{r} G^{(-1)} - \frac{1}{r} G^{(1)}] + Y_{jM,\alpha}^{(-1)} / 3 * [2\frac{\partial G^{(-1)}}{\partial r} + \frac{\sqrt{j(j+1)}}{r} G^{(1)} - \frac{2}{r} G^{(-1)}] \quad \text{(A4)}$$

The boundary conditions (7a) can be formulated as a continuity requirement on the value $\sigma_{\alpha\beta}n_\beta$

$$\sigma_{\alpha\beta} n_{\beta} = Y_{jM,\alpha}^{(1)} C_T^2 [\frac{\partial G^{(1)}}{\partial r} + \frac{\sqrt{j(j+1)}}{r} G^{(-1)} - \frac{1}{r} G^{(1)}] + Y_{jM,\alpha}^{(-1)} \{C_L^2 [\frac{\partial G^{(-1)}}{\partial r} - \frac{\sqrt{j(j+1)}}{r} G^{(1)} + \frac{2}{r} G^{(-1)}] + C_T^2 [\frac{2\sqrt{j(j+1)}}{r} G^{(1)} - \frac{4}{r} G^{(-1)}]\} \quad \text{(A5)}$$

For the transverse quasi polar (17b) and longitudinal (21) optic modes, the corresponding expressions are respectively:

$$G^{(1)} = G^{(1)}{}_T = [\frac{j}{2j+1} g_{j+1}(kr) + \frac{j+1}{2j+1} g_{j-1}(kr)],$$

$$G^{(-1)} = G^{(-1)}{}_T = \frac{\sqrt{j(j+1)}}{2j+1}[-g_{j+1}(kr) + g_{j-1}(kr)] \qquad (A6)$$

$$G^{(1)}=G^{(1)}{}_L=G_j \frac{\sqrt{j(j+1)}}{r},\ \ G^{(-1)}=G^{(-1)}{}_L=\frac{\partial G_j}{\partial r} \qquad (A7)$$

The functions $g_j(kr)$ are described in (16). The value of an ion plasma frequency (8b) is very large, $\Omega_p^2 >> \omega^2$, $\Omega_p^2 >> \omega_0(1,2)^2$. Therefore $\left|G^{(1)}\right| << \left|G^{(-1)}\right|$ and the LOW displacement is directed along the PNR radius when $\Omega_p \to \infty$. The functions $G_j$, $G_{j,in}$, $G_{j,out}$ can be written as follows:

$$\mathrm{r} < \mathrm{R},\ G_j = (2\pi)^{3/2} i^j J_{j+1/2}(K_L(2)r) / \sqrt{K_L(2)r},$$

$$\mathrm{r}>\mathrm{R},\ G_{j,in} = (2\pi)^{3/2} i^j H^{(2)}_{j+1/2}(K_L(2)r)/\sqrt{K_L(2)r},$$

$$\mathrm{r}>\mathrm{R},\ G_{j,out} = (2\pi)^{3/2} i^j H^{(1)}_{j+1/2}(K_L(2)r)/\sqrt{K_L(2)r},$$

$$K_L(1,2) = I\sqrt{\Omega_p^2+\omega_0^2(1,2)-\omega^2}/C_L \approx I\Omega_p / C_L,\ \left|K_L\right| R >> 1 \qquad (A8)$$

Finally, the full boundary conditions (A9a, b, c, d) contain the optic wave amplitudes $A_{T2}$, $A_{L2}$ (inside PNR) and the amplitudes $A_{T1in}$, $A_{T1out}$, $A_{L1out}$ (outside PNR). The momentum $K_L(1)$ is imaginary and we therefore have $A_{L1in}=0$.

$$A_{T2}G^{(1)}_{T2}+A_{L2}G^{(1)}_{L2}=A_{T1in}G^{(1)}_{T1in}+A_{T1out}G^{(1)}_{T1out} \qquad (A9a)$$

$$A_{T2}G^{(-1)}_{T2} + A_{L2}G^{(-1)}_{L2} = A_{T1in}G^{(-1)}_{T1in} + A_{T1out}G^{(-1)}_{T1out} + A_{L1out}G^{(-1)}_{L1out} \qquad (A9b)$$

$$A_{T2}\frac{\partial G^{(1)}_{T2}}{\partial r} + A_{L2}\frac{\partial G^{(1)}_{L2}}{\partial r} = A_{T1in}\frac{\partial G^{(1)}_{T1in}}{\partial r} + A_{T1out}\frac{\partial G^{(1)}_{T1out}}{\partial r} + A_{L1out}\frac{\partial G^{(1)}_{L1out}}{\partial r} \qquad (A9c)$$

$$A_{T2}\frac{\partial G^{(-1)}_{T2}}{\partial r} + A_{L2}\frac{\partial G^{(-1)}_{L2}}{\partial r} = A_{T1in}\frac{\partial G^{(-1)}_{T1in}}{\partial r} + A_{T1out}\frac{\partial G^{(-1)}_{T1out}}{\partial r} + A_{L1out}\frac{\partial G^{(-1)}_{L1out}}{\partial r} \qquad (A9d)$$

We are interested by the case when a value of LOW frequency $\Omega_p \to \infty$ and $|K_L.R|>>1$. Let us suppose that $A_{L2}G_{L2}^{(1)}$, $A_{L1out}G_{L1out}^{(1)} \sim 1/K_L$ so that we can omit terms $\sim A_{L2}$, $A_{L1out}$ in (A8a) (but ***not*** in (A8b, A8c)). We find expression (A10) from (A9b), (A9c) and (A7).

$$A_{T1in}\frac{\partial G^{(1)}_{T1in}}{\partial r} + A_{T1out}\frac{\partial G^{(1)}_{T1out}}{\partial r} - A_{T2}\frac{\partial G^{(1)}_{T2}}{\partial r} =$$
$$\frac{\sqrt{j(j+1)}}{r}\{A_{T1in}G^{(-1)}_{T1in} + A_{T1out}G^{(-1)}_{T1out} - A_{T2}G^{(-1)}_{T2}\} +$$
$$\frac{\sqrt{j(j+1)}}{r^2}\{A_{L1out}G_{jL1out} - A_{L2}G_{jL2}\} \qquad (A10)$$

We can omit terms ~ $A_{L1out}$ , $A_{L2}$, in (A10) and find the boundary conditions (26a), (26b) used in the main text of the paper.

The contribution of the LOW to the displacement parallel to PNR surface is small, ~$1/|K_L|$. However the contribution of the LOW to the displacement perpendicular to PNR surface is essential within a narrow layer of thickness ~$1/|K_L|$. This component of the LOW displacement leads to the existence of a non-zero term on the right hand side of Eq (26b).

In order to check the quality of our approximation $|K_L|R \to \infty$ we calculated the QP TOW scattering cross-section σ exactly (Eqs. A9a, b, c, and d) for the case $\Omega_p$=30 meV, *R*=6, *T*=130 K, $C_o$=70meV/r.l.u., $C_L$=105 meV/r.l.u. We found that this exact value of σ is in good agreement with the value of Σ (Fig. 5) calculated in the approximation $|K_L|R \to \infty$.

## ACKNOWLEDGMENTS

This work was partly supported by the US Department of Energy under grant DE-FG-06ER46318.